\documentclass[journal,10pt,twoside,a4paper]{IEEEtran}

\usepackage{xcolor,soul,framed} 

\colorlet{shadecolor}{yellow}
\usepackage[pdftex]{graphicx}
\graphicspath{{../pdf/}{../jpeg/}}
\DeclareGraphicsExtensions{.pdf,.jpeg,.png}

\usepackage[cmex10]{amsmath}
\usepackage{array, cite,xcolor}
\usepackage{mdwmath}
\usepackage{mdwtab}
\usepackage{eqparbox}
\usepackage{url}
\usepackage{makecell}

\usepackage{xcolor, multicol, tabularx,booktabs, amssymb, dsfont, multirow, cite}
\usepackage[utf8]{inputenc}
\usepackage{amsmath, cases, bbm, graphicx, amsthm, algpseudocode, algorithm, caption, subcaption}

\usepackage{balance}

\usepackage{caption}
\usepackage[hidelinks]{hyperref}
\usepackage{cleveref}

\usepackage{xcolor}
\usepackage{mathrsfs}
\usepackage{tikz}
\usetikzlibrary{calc}
\newcommand*\circled[1]{\tikz[baseline=(char.base)]{
    \node[shape=circle, draw, inner sep=1pt, 
        minimum height={\f@size*1.6},] (char) {\vphantom{WAH1g}#1};}}

\usepackage{hyperref}
\hypersetup{
 unicode=false,          
 pdftoolbar=true,        
 pdfmenubar=true,        
 pdffitwindow=false,     
 pdfstartview={FitH},    
 pdftitle={My title},    
 pdfauthor={Author},     
 pdfsubject={Subject},   
 pdfcreator={Creator},   
 pdfproducer={Producer}, 
 pdfkeywords={keyword1, key2, key3}, 
 pdfnewwindow=true,      
 colorlinks=true,       
 linkcolor=purple,          
 citecolor=purple,        
 filecolor=purple,      
 urlcolor=purple           
}

\usepackage{titlesec}
\titlespacing*{\section}{0pt}{0.5\baselineskip}{0.5\baselineskip}
\titlespacing*{\subsection}{0pt}{0.4\baselineskip}{0.4\baselineskip}

\makeatletter

\newcommand\sfcodefork{%
  \ifnum\the\spacefactor=1000 \expandafter\@firstoftwo\else\expandafter\@secondoftwo\fi
}

\makeatother
\usepackage{pifont}

\newtheorem{example}{Example}

\newtheorem{proposition}{Proposition}

\usepackage{pifont}
\newcommand{\cmark}{\ding{51}} 
\newcommand{\xmark}{\ding{55}} 

\crefname{lemma}{Lemma}{lemmas}

\newcolumntype{C}{>{\centering\arraybackslash}X} 
\newcolumntype{s}{>{\arraybackslash\hsize=.6\columnwidth}X}

\begin{document}

\markboth{IEEE Transactions on Intelligent Transportation Systems}%
{Nguyen \MakeLowercase{\textit{et al.}}: $\mathtt{Oranits}$: Mission Assignment and Task Offloading in Open RAN-based ITS}

\bstctlcite{IEEEexample:BSTcontrol}

\title{$\mathtt{Oranits}$: Mission Assignment and Task Offloading in Open RAN-based ITS using Metaheuristic and Deep Reinforcement Learning}

\author{
    Ngoc Hung~Nguyen, 
    Nguyen~Van Thieu,
    Quang-Trung~Luu,
    Anh Tuan~Nguyen,\\
    Senura Wanasekara,  
    Nguyen~Cong Luong,
    Fatemeh Kavehmadavani,
    and Van-Dinh~Nguyen
    
    \thanks{This work was supported in part by the Green Serverless Computing for Resource-Efficient AI Training Project at VinUniversity under Grant VUNI.CEI.FS 0002. Preliminary work was presented at  IEEE Global Communications Conference (GLOBECOM), Taipei, Taiwan, 8-12 Dec. 2025 \cite{nguyen2025metaheuristic}. 
    }

    \thanks{N. H. Nguyen, N. V. Thieu, and N. C. Luong are with the Phenikaa School of Computing, Phenikaa University, Duong Noi, Hanoi 12116, Vietnam (e-mail: \{hung.nguyenngoc, thieu.nguyenvan, luong.nguyencong\}@phenikaa-uni.edu.vn). \textit{Corresponding author: Ngoc Hung Nguyen.}}
    \thanks{Q. T. Luu is with Universit\'e Paris-Saclay - CNRS - CentraleSup\'elec - L2S, Gif-sur-Yvette, F-91192, France (email: quangtrung.luu@centrale supelec.fr).}
    \thanks{S. H. Wanasekara is with the School of Computer Science, The University of Sydney, Sydney, Australia (e-mail: wwan0281@uni.sydney.edu.au).}
    \thanks{A. T. Nguyen is with the Department of Smart City, Hanyang University, Ansan 15588, Korea (e-mail: natuan@hanyang.ac.kr).}
    \thanks{F. Kavehmadavan is with the Interdisciplinary Centre for Security, Reliability and Trust (SnT), University of Luxembourg (email: fatemeh.kavehmadavani@uni.lu).}
    \thanks{V.-D.~Nguyen is with the School of Computer Science and Statistics, Trinity College Dublin,  Dublin 2, D02 PN40, Ireland (e-mail: dinh.nguyen@tcd.ie), and was with the College of Engineering and Computer Science, VinUniversity, Hanoi 100000, Vietnam. The work of V.-D. Nguyen was done when he was with VinUniversity.}

    }
\maketitle

\begin{abstract}
    In this paper, we explore mission assignment and task offloading in an Open Radio Access Network (Open RAN)-based intelligent transportation system (ITS), where autonomous vehicles leverage mobile edge computing for efficient processing. Existing studies often overlook the intricate interdependencies between missions and the costs associated with offloading tasks to edge servers, leading to suboptimal decision-making. To bridge this gap, we introduce $\mathtt{Oranits}$, a novel system model that explicitly accounts for mission dependencies and offloading costs while optimizing performance through vehicle cooperation. To achieve this, we propose a twofold optimization approach. First, we develop a metaheuristic-based evolutionary computing algorithm, namely the Chaotic Gaussian-based Global ARO (CGG-ARO), serving as a baseline for one-slot optimization. Second, we design an enhanced reward-based deep reinforcement learning (DRL) framework, referred to as the Multi-agent Double Deep Q-Network (MA-DDQN), that integrates both multi-agent coordination and multi-action selection mechanisms, significantly reducing mission assignment time and improving adaptability over baseline methods.
    Extensive simulations reveal that CGG-ARO improves the number of completed missions and overall benefit by approximately $7.1\%$ and $7.7\%$, respectively. Meanwhile, MA-DDQN achieves even greater improvements of $11.0\%$ in terms of mission completions and $12.5\%$ in terms of the overall benefit. These results highlight the effectiveness of $\mathtt{Oranits}$ in enabling faster, more adaptive, and more efficient task processing in dynamic ITS environments.
\end{abstract}


\IEEEpeerreviewmaketitle
\section{Introduction}
\subsection{Background}
\IEEEPARstart{R}{ecent} advancements in smart city infrastructure and the growing demand for efficient transportation have accelerated the adoption of autonomous vehicles and the Internet of Things (IoT) \cite{10375912,9316663}. Driverless cars are revolutionizing transportation by enhancing safety, optimizing traffic flow, and improving overall efficiency \cite{10401029}. However, their seamless operation relies on real-time data processing, requiring substantial computational power and ultra-low latency to ensure reliability and responsiveness \cite{10384717, e0401697-b4a9-3a0f-8524-40c69e1a2891}. To meet these demands, autonomous vehicles leverage edge and cloud computing to efficiently handle complex tasks such as object detection, route planning, and real-time decision-making. These technologies facilitate seamless vehicle-to-vehicle and vehicle-to-infrastructure communication, enabling coordinated actions in dynamic environments \cite{BAKIRCI2024112015}. In addition, artificial intelligence (AI) is playing an increasingly vital role in intelligent transportation systems (ITS) by enhancing data accessibility and decision-making capabilities. However, this also leads to increased network congestion, particularly in dense urban areas and peak traffic hours \cite{9728752}. As a result, effective collaboration among autonomous vehicles and delivery robots is crucial to managing high demand, optimizing resource allocation, and preventing system overloads \cite{10097591}.

Mobile edge computing (MEC) has emerged as a key technology in modern computing architectures by bringing processing power closer to end-users. This proximity enables ultra-low latency, reduced energy consumption, and improved system efficiency \cite{9385940,10592058}. MEC servers support advanced algorithms, including machine learning (ML), deep reinforcement learning (DRL), and scheduling mechanisms, allowing them to handle diverse tasks in highly dynamic environments. As a result, MEC plays a pivotal role in ITS, ensuring seamless task execution, efficient resource utilization, and real-time decision-making across interconnected devices \cite{xu2022computation, li2024collaborative}.

Complementing MEC, the Open Radio Access Network (Open RAN) paradigm is transforming wireless communication by enhancing flexibility, interoperability, and scalability in radio access systems. By enabling operators to integrate components from multiple vendors, Open RAN reduces costs and fosters innovation, making it a cornerstone of next-generation mobile networks \cite{khan2024oran, li2025toward}. In ITS, Open RAN significantly reduces communication latency while improving AI model training and deployment. Leveraging open interfaces and standardized protocols, it simplifies network management and, when combined with virtualization and AI, optimizes operations such as dynamic bandwidth allocation and enhanced user experience \cite{ polese2023understanding}. These capabilities make Open RAN essential for ensuring seamless, low-latency communication between vehicles and infrastructure in an increasingly connected transportation landscape.

The integration of MEC and Open RAN further enhances their effectiveness in ITS by combining MEC’s high-speed, localized processing with Open RAN’s flexible and open architecture. This synergy offers several advantages: it enables infrastructure nodes to host both MEC servers and Open RAN components, consolidates monitoring databases for improved resource management, and facilitates cross-operations for enhanced coordination and functionality \cite{i2020perspective}. Together, MEC and Open RAN create a robust framework capable of meeting the stringent requirements of ITS, delivering low-latency processing, efficient resource allocation, and seamless AI deployment in highly dynamic environments.

\subsection{Research Gap, Motivation, and Contributions}
Currently, missions are often treated in isolation, overlooking their inherent interdependencies. This fragmented perspective can lead to system-wide inefficiencies, where improvements in one aspect may unintentionally introduce delays or resource conflicts in others. In reality, transportation tasks are interconnected and shared mobility services must synchronize passenger schedules while autonomous delivery fleets juggle routing, energy use, and computational constraints. With the rising demand for multi-purpose transportation, isolated optimization is no longer viable. The increasing reliance on MEC, AI, and real-time decision-making further complicates managing concurrent tasks. Without an integrated approach, network congestion, poor resource allocation, and computational overload can severely degrade system performance. Additionally, traditional wireless networks lack the flexibility needed to adapt to dynamic transportation environments.

Open RAN offers a transformative solution by enabling flexible, AI-driven resource allocation and seamless MEC integration, reducing latency and enhancing system coordination. To fully harness these advancements, we introduce $\mathtt{Oranits}$, a unified optimization framework that holistically manages mission interdependencies, optimizes resource efficiency, and ensures seamless coordination among autonomous vehicles, delivery robots, and smart infrastructure.

In summary, our contributions are as follows:
\begin{enumerate} 
    \item We propose $\mathtt{Oranits}$, a unified system that integrates Open RAN and MEC to optimize mission execution in ITS. $\mathtt{Oranits}$ addresses the complex interdependencies between missions, which are often overlooked in traditional scheduling, by considering offloading costs, processing locations, and mission execution order. It prioritizes high-impact missions to improve overall system efficiency. We also provide a detailed analysis of mission dependencies specific to ITS.               
    \item We formulate an optimization problem for mission scheduling and distribution in Open RAN-based ITS, incorporating constraints such as deadlines, routes, traffic, and network performance. For single-slot scheduling, we introduce the Chaotic Gaussian-based Global ARO (CGG-ARO), a metaheuristic algorithm that improves mission allocation, increases task completion and enhances system performance.
                      
    \item We extend the scheduling problem to dynamic and continuous-time scenarios where missions arrive periodically. To address this, we develop a DRL framework, namely the Multi-agent Double Deep Q-Network (MA-DDQN), that adapts in real time to real-time traffic conditions and environmental changes. Compared to CGG-ARO, the DRL model offers faster decision-making and better scalability for large-scale ITS deployments.
    
    \item  We validate our methods through comprehensive benchmarking against state-of-the-art (SOTA) metaheuristics. Results show that our approach improves system profit and mission completion rates, consistently outperforming existing solutions.          
\end{enumerate}

\subsection{Mathematical Notations and Paper Organization}
\begin{table}[t]
    \centering
    \caption{Mathematical Notations.}
    \renewcommand{\arraystretch}{1.2} 
    \setlength{\tabcolsep}{10pt}
    \begin{tabularx}{\linewidth}{c X} 
        \hline
        \textbf{Notation} & \textbf{Description} \\
        \hline
        $K$ and $K^*$ & Total number of vehicles in the system and the subset of vehicles assigned to a specific solution row\\
        $\mathcal{S}$ & Set of servers, including MEC servers $\mathcal{S}^{m}$ and one cloud server\\
        $\tau$ & Time constraint for completing all missions\\
        $M(\tau)$ & Total number of missions within time $\tau$\\
        $Z$ & Number of missions in a specific solution row\\
        $N$ & Number of mission groups\\
        $M_i(\tau)$ & The $i$-th mission in a specific row\\
        $\theta_{M_i(\tau)}$& Vehicle assigned to handle mission $M_i(\tau)$\\
        $\mathcal{J}_{i}$ & Set of offloading tasks associated with mission $M_i(\tau)$\\
        $U$ & Number of uplink channels used for communication between vehicles and a RU\\
        $W_{c}$ & Communication bandwidth for one channel\\
        $\mathds{1}_{\{\cdot\}}$ & Indicator function. \\
        \hline
    \end{tabularx}
    \label{tab:notations}
\end{table}
The remainder of this paper is organized as follows. Section~\ref{sec:relatedwork} states the related work. Section~\ref{sec:sys_overview} introduces the system model, mission organization, and Open RAN architecture in ITS. Section~\ref{sec:problem_formulation} mathematically formulates the mission assignment and task offloading problem in Open RAN-based ITS. Section~\ref{sec:Metaheuristic-Approach} presents our metaheuristic algorithm and CGG-ARO based on evolutionary computing, while Section~\ref{sec:DRL-Approach} introduces a DRL-based approach. Section~\ref{sec:numerical_simulation} provides a detailed analysis of the simulation results, followed by conclusions and future research directions in Section~\ref{sec:cons}.

To improve readability, the main mathematical notations used throughout the paper are summarized in \autoref{tab:notations}.

\begin{table*}[!t]
    \centering
    \caption{Comparison of Representative Related Works in Open RAN-based ITS.}
    \resizebox{\textwidth}{!}{%
    \renewcommand{\arraystretch}{0.9} 
    \setlength{\tabcolsep}{7pt}
    \begin{tabular}{l c c c c c c c}
        \toprule
        \text{Reference} 
        & \makecell{\text{Mission} \\ \text{interdependence}} 
        & \text{Metaheuristic} 
        & \text{DRL/AI} 
        & \text{Open RAN} 
        & \makecell{\text{Offloading} \\ \text{cost}} 
        & \makecell{\text{MEC-task} \\ \text{integration}} 
        & \makecell{\text{Multi-agent} \\ \text{framework}} \\
        \midrule
        \cite{li2024collaborative} & \xmark & \xmark & \cmark & \xmark & \cmark & \cmark & \cmark \\
        \cite{li2025toward} & \xmark & \xmark & \cmark & \cmark & \xmark & \cmark & \xmark \\
        \cite{sartori2022scheduling} & \cmark & \xmark & \xmark & \xmark & \xmark & \xmark & \xmark \\
        \cite{liu2022iterative} & \cmark & \xmark & \xmark & \xmark & \xmark & \xmark & \xmark \\
        \cite{ding2024intelligent} & \xmark & \cmark & \xmark & \xmark & \xmark & \xmark & \xmark \\
        \cite{moshiri2025joint} & \xmark & \cmark & \xmark & \xmark & \cmark & \cmark & \xmark \\
        \cite{sroka2024policy} & \xmark & \xmark & \xmark & \cmark & \xmark & \xmark & \xmark \\
        \cite{seid2025multi} & \xmark & \xmark & \cmark & \cmark & \xmark & \xmark & \cmark \\
        \midrule
        \textbf{Our work} & \cmark & \cmark & \cmark & \cmark & \cmark & \cmark & \cmark \\

        \bottomrule
    \end{tabular}%
    }
\label{tab:related_work_comparison}
\end{table*}

\section{Related Work}
\label{sec:relatedwork}
Processing missions such as passenger transport, food delivery, and goods shipping are central to ITS, alongside optimizing traffic flow, enhancing safety, and improving efficiency \cite{wang2019demystifying,del2019bio}. These missions often exhibit interdependencies with other jobs or routes \cite{10247090,8886346, sartori2022scheduling}. For instance, if one passenger group depends on another, vehicles must transport them sequentially or collaborate to minimize delays. Such dependencies impact routing efficiency and overall system performance \cite{FantinIrudayaRaj2022,liu2022iterative}.

 Metaheuristics provide powerful optimization methods by enabling randomized search processes \cite{satouf2025metaheuristic}. 
Among them, genetic algorithms (GA) leverage principles of biological evolution to solve complex problems. Similar nature-inspired techniques include particle swarm optimization (PSO) \cite{moshiri2025joint}, artificial rabbits optimization (ARO) \cite{wang2022artificial}, success-history adaptation differential evolution (SHADE) \cite{kaveh2024success}, and linear population size reduction SHADE (L-SHADE) \cite{ghosh2022using}. These algorithms efficiently balance exploration and exploitation, with PSO simulating swarm behavior, ARO mimicking rabbit hunting strategies, and SHADE dynamically adjusting parameters for faster convergence \cite{van2023mealpy, goudos2019artificial}.
Despite their effectiveness, their long search times result from exploring vast solution spaces, leading to high computational costs \cite{kaveh2023application}. Additionally, many rely on fixed parameters such as population size and mutation rates, which can limit adaptability to dynamic environments and reduce search efficiency \cite{omidvar2021review}.

Recently, ML and deep learning (DL) have been widely applied across various domains, including computer vision, natural language processing, and generative AI. These AI-driven approaches have also gained traction in optimization tasks, particularly through deep reinforcement learning (DRL), imitation learning, and other DL-based techniques \cite{yan2023edge, mashal2024multiobjective, agarwal2025open, ding2024intelligent}. Among these, DRL has emerged as a key solution for optimizing intelligent networks, enabling efficient resource allocation and accurate channel estimation \cite{yan2026dynamic, vidya2025dynamic}.
In particular, \cite{li2024collaborative} employs multi-agent DRL to jointly address task offloading and resource allocation in small-cell MEC systems. The proposed approach leverages the centralized training distributed execution framework to enable efficient distributed decision-making, achieving near-centralized performance while effectively reducing total energy consumption under delay constraints. Similarly, \cite{li2025toward} proposes a unified DRL deployment architecture, where DRL-based solutions can be applied in real-world scenarios with the support of Open RAN and MEC, ensuring a consistent and practical interface. In addition, DRL combined with federated learning has attracted increasing attention due to its ability to enhance privacy while maintaining strong system performance \cite{abou2024federated}. 
Overall, AI-driven approaches have significantly improved Open RAN systems by optimizing network management, automating resource scheduling, and enhancing adaptability to dynamic environments \cite{nguyen2023network, kavehmadavani2024empowering}. With ongoing advancements in DL architectures and training strategies, DRL has evolved into various frameworks, including single-agent and multi-agent systems, value-based and policy-based methods, and multi-action models \cite{yang2024multiagent, wang2024deep}. These developments further enable more scalable, adaptive, and autonomous network optimization.

To further clarify the capabilities and limitations of the literature most relevant to our problem, we present a comparative overview in \autoref{tab:related_work_comparison}. The table combines two representative ITS scheduling studies that explicitly capture mission interdependence \cite{sartori2022scheduling, liu2022iterative} with more recent works on metaheuristic optimization \cite{ding2024intelligent, moshiri2025joint}, multi-agent DRL-based MEC offloading \cite{li2024collaborative}, and AI-enabled Open RAN control \cite{sroka2024policy, seid2025multi, li2025toward}. As can be seen, the existing literature typically emphasizes only one or two aspects of the overall problem at a time. Mission interdependence is mainly addressed in transportation-oriented scheduling studies, while recent AI and Open RAN works focus more on automation, architectural deployment, or resource control. Only a limited subset jointly considers offloading-aware decision-making, MEC integration, Open RAN support, and scalable multi-agent coordination. These gaps highlight the need for a unified framework that can integrate mission-aware routing, metaheuristic search, AI-enhanced decision-making, and architectural flexibility, which is the main contribution of our work.

\section{System Model}
\label{sec:sys_overview}
\subsection{System Overview}
\label{sec:problem_descriptions}
\begin{figure}[t]
    \centering	
    \includegraphics[width=0.9\columnwidth]{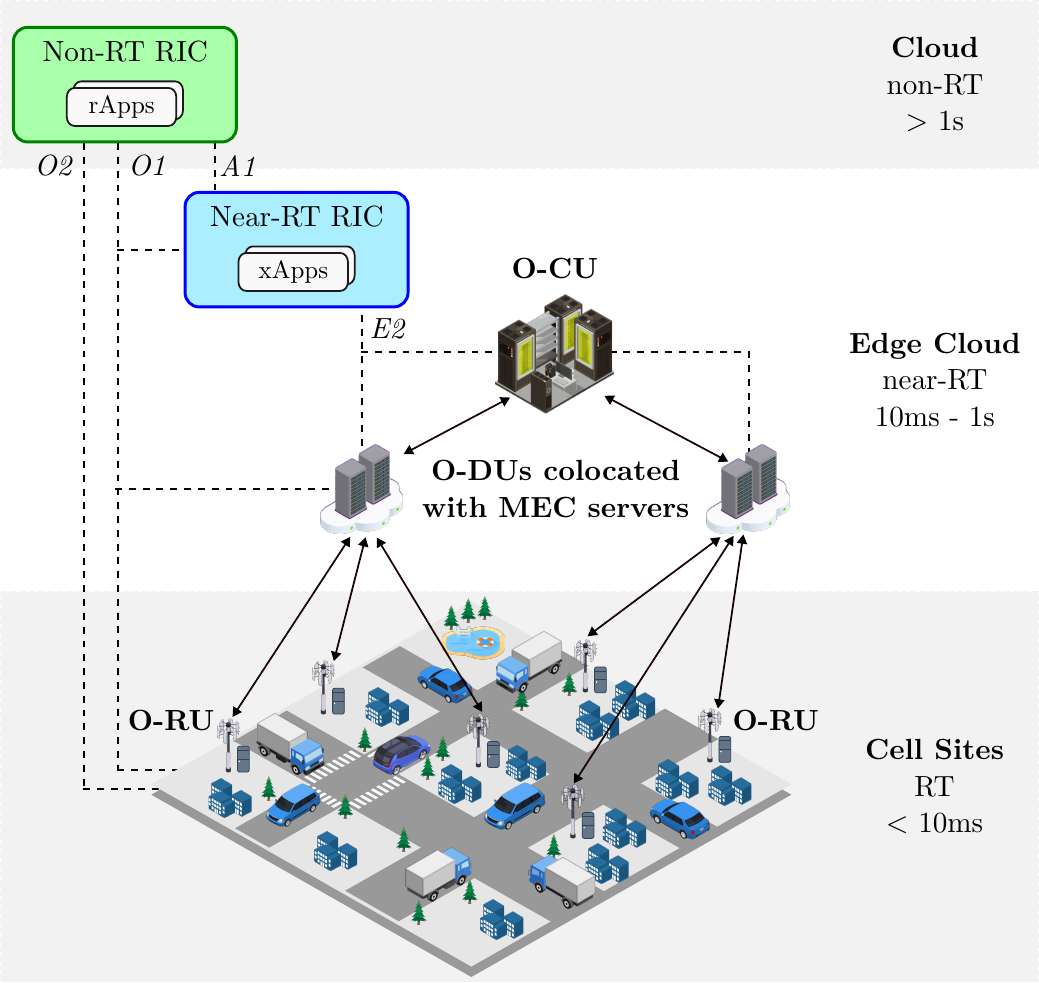}
    \caption{$\mathtt{Oranits}$: Integration of Open RAN and MEC to enable ITS.}
    \label{fig_ORAN-ITS}
\end{figure}

Fig.~\ref{fig_ORAN-ITS} illustrates the $\mathtt{Oranits}$ system model, which is the integration of Open RAN and MEC to support ITS. The architecture consists of a set of servers $\mathcal{S}$, categorized into one cloud server $\mathcal{S}^{c}$ and MEC servers $\mathcal{S}^{m}$. In this model, distributed units (DUs) are deployed at MEC servers to provide localized computational capabilities, while centralized units (CUs) reside in the cloud to handle high-performance computing tasks. Each DU is linked to a radio unit (RU), which serves as an access point for autonomous vehicles operating in the system. The Open RAN architecture consists of three layers: management, control, and function. The management layer placed in the cloud operates in non-real-time ($>1$s) for orchestration, automation, and AI/ML model deployment. The control layer placed at the edge cloud works in near-real-time ($10$ms to $1$s), handling radio resource management, quality of service (QoS), and interference management. The function RAN layer runs below $10$ms for tasks like scheduling and power control.
Open RAN introduces the non-real-time RAN intelligent controller (Non-RT RIC) and Near-RT RIC. The Non-RT RIC supports AI/ML workflows and policy guidance, while the Near-RT RIC handles real-time RAN control and optimization \cite{kavehmadavani2024empowering}.

Over a given time period $\tau$ (\textit{i.e.}, from $t$ to $t + \tau$), let $\mathcal{K}(\tau)$ denote the set of $K$ available vehicles and $M(\tau)$ represent the number of arriving missions. To simplify scheduling, missions are grouped into $N = \lceil M(\tau) / Z \rceil$ subsets of size $Z$. If $M(\tau) \mod Z \neq 0$, the last subset is padded with empty missions to maintain uniformity. The mission set is then structured as a matrix $\mathbf{M}(\tau) \in \mathbb{R}^{N \times Z}$:
\begin{equation}
    \mathbf{M}(\tau) =
    \begin{bmatrix}
        M_{1,1}(\tau) & M_{1,2}(\tau) & \cdots & M_{1,Z}(\tau) \\
        M_{2,1}(\tau) & M_{2,2}(\tau) & \cdots & M_{2,Z}(\tau) \\
        \vdots & \vdots & \ddots & \vdots \\
        M_{N,1}(\tau) & M_{N,2}(\tau) & \cdots & M_{N,Z}(\tau)
    \end{bmatrix}_{N \times Z}
\end{equation}
wherein each element $M_{n,i}(\tau)$ represents the $i$-th mission in the $n$-th subset, for all $i \in [1,Z]$ and $n \in [1,N]$. In the following, mission assignment optimization is performed on each mission subset (\textit{i.e.}, each row $n$ of $\mathbf{M}(\tau)$). To simplify notation, we denote $\mathcal{M}(\tau)$ as the selected row in $\mathbf{M}(\tau)$ and replace $M_{n,i}(\tau)$ with $M_{i}(\tau)$ to represent the $i$-th mission within a specific subset $n$.

\textbf{Mission description}:
A given mission  $M_{i}(\tau)$ is represented by a tuple  
\begin{equation} \label{eq:MissionDescription}
    M_{i}(\tau) \triangleq 
    \big\langle
        r_i(\tau), 
        T_i(\tau),
        B_i(\tau), 
        \mathcal{M}_{i}^{-}, 
        \mathcal{M}_{i}^{+}
    \big\rangle 
\end{equation}
where $r_i(\tau)$ represents the route information for completing mission $M_{i}(\tau)$, including the start point $r_i^{\text{start}}(\tau)$ and the end point $r_i^{\text{end}}(\tau)$,
$T_i(\tau)$ is the deadline for mission completion,
$B_i(\tau)$ denotes the allocated budget for offloading costs to cloud/edge servers,
$\mathcal{M}_{i}^{-} \subset \mathcal{M}\left(\tau\right)$ is the set of predecessor missions that must be completed before $M_{i}(\tau)$, and 
$\mathcal{M}_{i}^{+} \subset \mathcal{M} \left(\tau\right)$ 
is the set of successor missions that depend on the completion of $M_{i}(\tau)$.
If $\mathcal{M}_{i}^{-} = \varnothing$, the mission can start immediately without delay. Similarly, if $\mathcal{M}_{i}^{+} = \varnothing$, the completion of $M_i(\tau)$ does not impact the execution of other missions.

\textbf{Tasks offloading}:
While executing missions on the road, vehicles must process autonomous tasks, manage in-vehicle entertainment systems, and handle various functions that enhance user experience. The execution of these tasks is influenced by road conditions, computational capacity, vehicle speed, and occasionally, user-initiated on-demand requests. Given the continuous environmental sensing required for navigation, vehicles must process an overwhelming volume of autonomous tasks \cite{9773158}. To improve efficiency, edge AI systems are designed to handle these tasks in real time. By offloading critical data (\textit{e.g.}, speed, road identifiers, and task dependencies) to edge or cloud servers, vehicles can significantly reduce processing time, minimize energy consumption, and ensure ultra-low latency \cite{8744265, 9430907}. Additionally, offloading tasks are independent of one another, allowing them to be transmitted immediately upon the vehicle's decision to offload. While these tasks can be processed locally, limited computational resources and storage capacity may lead to system overload if the vehicle attempts to handle all tasks on its own.

Let $\mathcal{J}_{i}$ represent the set of tasks that a vehicle handling mission $M_i(\tau)$ must offload during its operation. Each task $j \in \mathcal{J}_{i}$ is defined as a tuple:
$\langle \alpha_{i,j}(\tau), \beta_{i,j}(\tau) \rangle$, 
where $\alpha_{i,j}(\tau)$ denotes the input transmission size (in bits) and
$\beta_{i,j}(\tau)$ represents the number of CPU cycles required to process task $j$.
We further assume that the feedback data size from the edge/cloud server to vehicles is negligible compared to the input data size. 
\subsection{Traffic Status}
Traffic conditions are classified into five distinct categories, which remain unchanged throughout the time duration $\tau$:
1) \textbf{Free flow:} Vehicles travel at maximum speed without encountering obstacles. 2) \textbf{Stable flow:} Vehicles move steadily, with a slight reduction in speed due to increased vehicle density. While no congestion occurs, minor delays might arise from traffic lights or yielding to other vehicles. 3) \textbf{Slow flow:} Increased vehicle density leads to significant speed reductions. Vehicles may frequently slow down or make occasional stops, but traffic remains in motion. 4) \textbf{Congested flow:} Heavy traffic results in slow movement or complete stops, causing substantial delays even for short distances. 5) \textbf{Severe congestion:} Traffic is nearly at a standstill, with vehicles either stationary or moving at extremely slow speeds. This situation typically arises during peak hours or due to accidents.

Each road segment is assigned a traffic status coefficient $c_i$ (where $i$ corresponds to the road index), representing its congestion level \cite{fan2026task}. Based on this coefficient, the shortest paths for mission execution and vehicle routing are determined using Dijkstra's algorithm \cite{noto2000method}.

\section{Problem Formulation}
\label{sec:problem_formulation}
\subsection{Mission Assignment}
\begin{figure}[t]
    \centering
    \includegraphics[width=0.95\columnwidth]{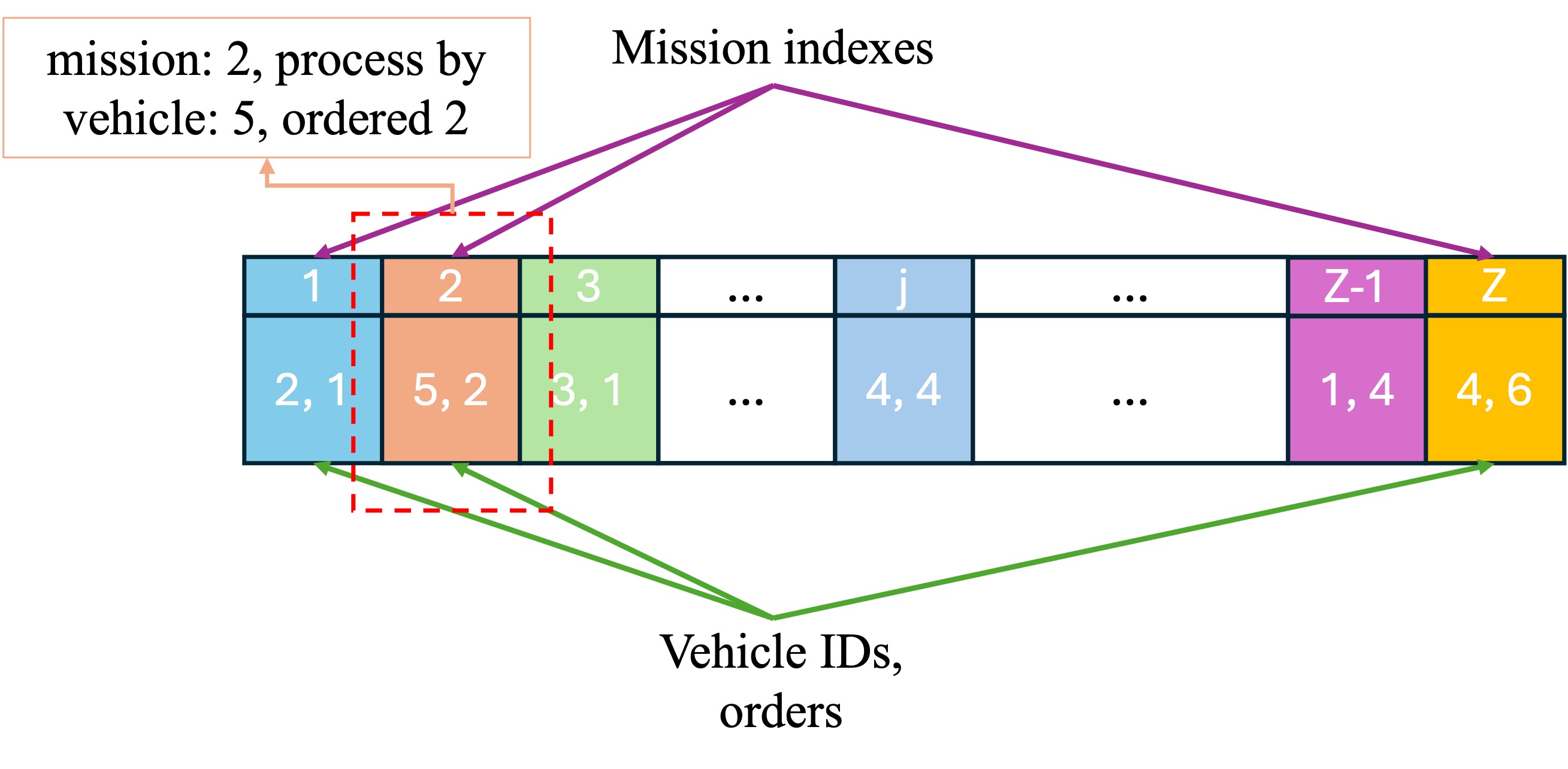}
    \caption{Example of assigning $Z$ missions to five vehicles.}
    \label{fig:MissionAssignmentExample}
\end{figure}
Let $\mathbf{D}(\tau)$ represent the solution for assigning $N$ rows of missions from the mission set $\mathbf{M}(\tau)$. Each row $\mathbf{M}_{n,:}(\tau)$ contains $Z$ missions, whose assignments to available vehicles are denoted by $\mathbf{D}_{n,:}(\tau)$ (\textit{i.e.}, row $n$ of $\mathbf{D}(\tau)$). Each element $D_{n,i}(\tau)$ in $\mathbf{D}_{n,:}(\tau)$ is a tuple $\langle \theta_{M_{i}(\tau)}, \sigma_{M_{i}(\tau)} \rangle$, where $\theta_{M_{i}(\tau)}$ denotes the vehicle assigned to mission $M_i(\tau)$, and $\sigma_{M_{i}(\tau)}$ represents its processing order. Thus, the assignment matrix is given by:
\begin{equation}
    \mathbf{D}_{n,:}(\tau) = 
    \big
    [
        \langle
            \theta_{M_{i}(\tau)}, \sigma_{M_{i}(\tau)}
        \rangle,
        i \in [1,Z]
    \big].
\end{equation}

\begin{example} \label{ex:1}
    \autoref{fig:MissionAssignmentExample} illustrates an assignment of $Z$ missions in row $n$ to five vehicles. In this example, mission $2$ is assigned to vehicle $5$ with an execution order of $2$, while missions $1$ and $3$ are handled by vehicles $2$ and $3$, both executed as the first task.
    The corresponding solution matrix $\mathbf{D}_{n,:}(\tau)$ is
    \begin{equation}
        \mathbf{D}_{n:}(\tau) = 
        \left[\begin{array}{c}
            \left\langle 2,1\right\rangle \\
            \left\langle 5,2\right\rangle \\
            \left\langle 3,1\right\rangle \\
            \vdots\\
            \left\langle 4,6\right\rangle 
            \end{array}\right]^{\top}\begin{array}{c}
            \text{mission }1,\\
            \text{mission }2,\\
            \text{mission }3,\\
            \vdots\\
            \text{mission }Z.
        \end{array}
    \end{equation}
\end{example}

In this paper, we assume that each subset $\mathbf{M}_{n,:}(\tau)$ of $Z$ missions is assigned to a fixed number of $K^* \leq |\mathcal{K}(\tau)|$ available vehicles, selected based on their proximity to the mission locations. This constraint is expressed as:
\begin{equation} \label{cons:MissionVehicleLimit}
    \sum_{M_{i}\left(\tau\right)\in\mathbf{M}_{n,:}\left(\tau\right)}\sum_{k\in\mathcal{K}\left(\tau\right)}\mathds{1}_{\left\{ k=\theta_{M_{i}(\tau)}\right\} }=K^{*}, ~\forall n\in\left[1,N\right].
\end{equation}
Each vehicle is therefore assigned approximately $\left\lceil Z/K^* \right\rceil$  missions. The $K^*$ selected vehicles are determined by an xApp implemented in the Near-RT RIC of the Open RAN system, ensuring minimal travel time to the assigned mission locations.

From the tuple $\langle     
    \theta_{M_{i}(\tau)}, \sigma_{M_{i}(\tau)}
\rangle$
of each element $D_{n,i}(\tau)$ of the solution matrix $\mathbf{D}(\tau)$, if
$\theta_{M_{i}(\tau)} = \theta_{M_{i'}(\tau)}$ ($\forall i,i' \in [1,Z]: i \neq i'$), that means both mission $M_{i}(\tau)$ and $M_{i'}(\tau)$ are handled by the same vehicle, and vice versa.
One can also deduce the mission scheduling order $\sigma_{k}$ of each vehicle $k \in \mathcal{K}(\tau)$ as
\begin{equation}
    \sigma_{k} = 
    \left\{ 
        \begin{array}{l}
            \forall i\in[1,Z]: \theta_{M_{i}(\tau)} = k,\\
            \text{sorted by }\sigma_{M_{i}(\tau)} \text{ in ascending order}
        \end{array}
    \right\}.
\end{equation}

A solution $\mathbf{D}(\tau)$ is valid only if it satisfies the following constraints.
First, each mission $M_{i}(\tau)$ must be assigned to exactly one vehicle:
\begin{align} \label{cons:MissionVehicle}
    \sum_{k\in\mathcal{K}\left(\tau\right)}\mathds{1}_{\left\{ k = \theta_{M_{i}(\tau)}\right\} } = 1, \quad \forall M_{i}\left(\tau\right)\in\mathbf{M}\left(\tau\right).
\end{align}
Next, the constraint below guarantees that each mission can be assigned at most one scheduling order:
\begin{align} \label{cons:MissionOnlyOneOrder}
     \sum_{\sigma=1}^{\left|\sigma_{k}\right|}\mathds{1}_{\left\{ \sigma_{M_{i}\left(\tau\right)}=\sigma\right\} } \leq 1, \quad \forall k\in\mathcal{K}\left(\tau\right), M_{i}\left(\tau\right)\in\mathbf{M}\left(\tau\right).
\end{align}

In addition, missions assigned to the same vehicle must have distinct scheduling orders:
{\small\begin{align} \label{cons:MissionOrderUnique}
    \sigma_{M_{i}\left(\tau\right)} \neq \sigma_{M_{i'}\left(\tau\right)}, 
       \forall M_{i}(\tau), M_{i'}(\tau) \in \mathbf{M}(\tau): \theta_{M_{i}(\tau)} = \theta_{M_{i'}(\tau)}.
\end{align}}Furthermore, for any given mission $M_{i}(\tau)$, its predecessor missions $M_{i'}(\tau) \in \mathcal{M}_{i}^{-}$ must have a lower scheduling order, while its successor missions $M_{i'}(\tau) \in \mathcal{M}_{i}^{+}$ must have a higher scheduling order:
\begin{subequations} \label{cons:MissionOrderBeforeAfter}
    \begin{align} 
        & \sigma_{M_{i'}\left(\tau\right)} < \sigma_{M_{i}\left(\tau\right)},\quad\forall M_{i'}\left(\tau\right)\in\mathcal{M}_{i}^{-}  \\
        & \sigma_{M_{i'}\left(\tau\right)} > \sigma_{M_{i}\left(\tau\right)},\quad\forall M_{i'}\left(\tau\right)\in\mathcal{M}_{i}^{+}.
    \end{align}
\end{subequations}

\subsection{Offloading Strategy}
We assume that a vehicle $k \in \mathcal{K}(\tau)$ can seamlessly communicate with any MEC server within its coverage radius $R_k$. If the vehicle moves beyond this range and needs to offload a task, it must rely on a cloud server to maintain optimal latency performance. This approach ensures efficient task offloading, minimizes delays, and enhances overall system responsiveness.

In this paper, we implement a greedy offloading policy for all vehicles, which follows these steps: ($i$) The vehicle sends a request to all MEC servers within its coverage radius, querying available transmission and computing resources; $ii$) It estimates the offloading latency for each server; $iii$) The server with the lowest latency is selected, and ($iv$) 
The vehicle compares the cost and latency of the selected MEC server with the cloud server and chooses the option that minimizes latency.
Under this strategy, the offloading decision for a task is made at the time of upload, meaning the system assigns the vehicle to a specific MEC or cloud server beforehand. Specifically, let $S_o \in \mathcal{S}$, where $o \in [1,|\mathcal{S}|]$, denote the server handling a given offloading task.

\subsection{Mission Completion Time}
In this paper, we assume that all vehicles cooperate to complete missions on time. Each vehicle must reach the starting point of its mission route, $r_i^{\text{start}}(\tau)$, punctually to avoid cascading delays in subsequent tasks. Efficient task distribution and processing are therefore critical. For instance, when a vehicle encounters an obstacle, it must collect essential data, such as images, sensor readings, and navigation details, before deciding on the next action. However, limited onboard resources and road conditions can slow this process. To mitigate these challenges, vehicles can offload tasks to edge or cloud servers, though this introduces additional delays primarily influenced by communication time and computation time, as detailed below.

\textbf{Communication delay}:  When a vehicle decides to offload a task to the cloud server, it first transmits the task to a nearby AP, which then forwards it to the cloud. The fiber optic propagation delay for task $j$ of mission $M_i(\tau)$ is given by:
\begin{align}
    d^{\mathtt{fib}}_{i,j} = \frac{\alpha_{i,j}(\tau)}{R^{\mathtt{fib}}}
\end{align}where $R^{\mathtt{fib}}$  is the fiber optic transmission rate (bps). Next, we assume that each RU is equipped with $E$ antennas, supporting $U$ uplink communication channels via frequency division multiple access (FDMA). The bandwidth allocated to each RU is denoted by $W$ (Hz). Under FDMA, the bandwidth of each channel at an RU is given as $W_c=W/U$.
For vehicle $k = \theta_{M_i(\tau)}$ assigned to mission $M_i(\tau)$, let $\mathbf{h}^{k, o}_{i,j}(t) \in \mathbb{C}^{E \times 1}$ be the channel vector for task $j$ during the execution of mission $i$ from vehicle $k\in\mathcal{K}(\tau)$ to server $o\in\mathcal{S}
$. The throughput $R_{i,j}^{k,o}(t)$ (bps) for task $j$ is then computed as
\begin{align} \label{eq:rate}
R_{i,j}^{k,o}(t)=  W_c\log\left(1+\frac{p_k\|\mathbf{h}^{k, o}_{i,j}(t)\|^2}{ W_c N_0}\right)
\end{align}
where $p_{k}(t)$  is the transmission power of vehicle $k$ and $N_0$  is the noise power spectral density. 

Let $ d^{\text{comm}}_{i,j}(\tau)$ denote the communication delay experienced by vehicle $k$ when transmitting an offloading task to a RU. This occurs when the vehicle requests task $j \in \mathcal{J}_{i}$ to be processed by either an MEC or a cloud server. The communication delay is given by: 
\begin{align} d^{\text{comm}}_{i,j}(\tau) = \begin{cases} \frac{\alpha_{i,j}(\tau)}{R^{k,o}_{i,j}(\tau)}, & \text{if } S_o \in \mathcal{S}^m;\\ 
\frac{\alpha_{i,j}(\tau)}{R^{k,o'}_{i,j}(\tau)} + d^{\mathtt{fib}}_{i,j}, & \text{otherwise}\end{cases} 
\end{align} 
where $R^{k,o'}_{i,j}(\tau)$ represents the transmission rate from vehicle $k$ to server $o'\in\mathcal{S}
$, and $d^{\mathtt{fib}}_{i,j}$ accounts for additional fiber-optic transmission delay when communicating with a cloud server. The total communication delay experienced by vehicle $k$ for offloading all tasks $j \in \mathcal{J}_{i}$ while executing mission $M_i(\tau)$ is given by:
\begin{equation}
    d^{\text{comm}}_{i}(\tau) = \sum_{j=1}^{|\mathcal{J}_{i}|} d^{\text{comm}}_{i,j}(\tau).
\end{equation}

\textbf{Computation delay}:
The computation delay experienced by vehicle $\theta_{M_{i}(\tau)}$ while executing mission $M_i(\tau)$ can be expressed by 
\begin{align}
    d^{\text{comp}}_{i,j}(\tau) = \frac{\beta_{i,j}(\tau)}{f_{o}(t)}, \quad \forall t \in \tau
\end{align}
where $f_{\text{o}}(t)$ is the computational capacity of an assigned MEC or cloud server at time instant $t$ within period $\tau$.
The total computation delay of vehicle $\theta_{M_{i}(\tau)}$ for offloading all tasks $j \in \mathcal{J}_{i}$  while handling mission $M_i(\tau)$ is thus 
\begin{equation}
    d^{\text{comp}}_{i}(\tau) = \sum_{j=1}^{|\mathcal{J}_{i}|} d^{\text{comp}}_{i,j}(\tau).
\end{equation}

\textbf{Travel delay}:
For vehicle $\theta_{M_{i}(\tau)}$, we let $\bar{v}_i$ denote its ideal average speed, assuming no offloading time is considered. In reality, the vehicle’s effective speed on the road is heavily influenced by the offloading process, which depends on both its frequency and efficiency. The ideal travel delay $d^{\text{move}}_i$ for vehicle $\theta_{M_{i}(\tau)}$ on route $r_{i}(\tau)$ to complete mission $M_i(\tau)$ is computed as:
\begin{align}
    d^{\mathrm{move}}_{i}(\tau) = \frac{|r_{i}(\tau)|}{\bar{v}_i}
\end{align}
 where $|r_{i}(\tau)|$ denotes the length (in meters) of route $r_{i}(\tau)$, and $\bar{v}_i$ represents the average speed of vehicle $i$ (in meters per second). Note that both $|\mathcal{J}_i|$ and $\bar{v}_i$ depend on traffic conditions $\bar{c}_i$ along  $r_i(\tau)$, where $\bar{c}_i$ is obtained by aggregating the traffic states $c_l$, $l \in r_i(\tau)$, of all road segments along the route \cite{fan2026task}.

\textbf{The total delay}:
We calculate the total delay $d_i(\tau)$ that vehicle $\theta_{M_{i}(\tau)}$ experienced by vehicle $M_i(\tau)$ \textit{in isolation} (\textit{i.e.} without considering interdependent or 
pre-scheduled missions). To simplify the model, we assume no overlap between travel time $d^{\text{move}}_i$,  communication time $d^{\text{comm}}_i$, and computation time $d^{\text{comp}}_i$. The  total delay is thus given by:
\begin{align} \label{eq:MCT_isolation}
d_{i}(\tau) = 
d^{\text{move}}_{i}(\tau) 
+ d^{\text{comm}}_{i}(\tau) 
+ d^{\text{comp}}_{i}(\tau).
\end{align}

\textbf{Mission completion time}:
The overall mission completion time (MCT) for each mission $M_{i}(\tau)$ accounts for the total delay $d_i(\tau)$ incurred while executing $M_i(\tau)$, the completion time of prior-scheduled missions $M_{i'}(\tau)$ ($i' \neq i$) assigned to the same vehicle, and the completion time of interdependent missions  $M_{i'}(\tau) \in \mathcal{M}_{i}(\tau)$ executed by other vehicles.

Given the mission scheduling orders $\sigma_{k}$ at all vehicles $k \in \mathcal{K}(\tau)$, we derive a lower bound on the overall MCT $\delta_{i}(\tau)$ for mission $M_{i}(\tau)$ as follows:
\begin{equation} \label{cons:MissionCompletionTimeBound}
    \delta_{i}(\tau)
    \geq d_{i}\left(\tau\right) 
    + \sum_{i'\in\mathscr{I}_{1}}d_{i'}\left(\tau\right)
    + \sum_{i'\in\mathscr{I}_{2}}\left[
        d_{i'}\left(\tau\right)
        + \sum_{i''\in\mathscr{I}_{3}}d_{i''}\left(\tau\right)
    \right]
\end{equation}
where the second term $\sum_{i'\in\mathscr{I}_{1}}d_{i'}\left(\tau\right)$ accounts for the total time required to complete all missions $M_{i'}(\tau)$ scheduled before $M_{i}(\tau)$ on the same vehicle, and the third term represents the time required to complete all missions $M_{i'}(\tau)$ assigned to other vehicles that belong to the predecessor mission set $\mathcal{M}_{i}^{-}$ of mission $M_{i}(\tau)$. The sets $\mathscr{I}_{1}$, $\mathscr{I}_{2}$, and $\mathscr{I}_{3}$ are defined as follows:
\begin{align}
    \mathscr{I}_{1} & \triangleq\left\{ \forall i'\neq i:\theta_{M_{i'}\left(\tau\right)}=\theta_{M_{i}\left(\tau\right)},\sigma_{M_{i'}\left(\tau\right)}<\sigma_{M_{i}\left(\tau\right)}\right\} \\
    \mathscr{I}_{2} & \triangleq\left\{ \forall i'\neq i:\theta_{M_{i'}\left(\tau\right)}\neq\theta_{M_{i}\left(\tau\right)},M_{i'}\left(\tau\right)\in\mathcal{M}_{i}^{-}\right\} \\
    \mathscr{I}_{3} & \triangleq\left\{ \forall i''\neq i':\theta_{M_{i''}(\tau)}=\theta_{M_{i'}(\tau)},\sigma_{M_{i''}(\tau)}<\sigma_{M_{i'}(\tau)}\right\}.
\end{align}
\subsection{Mission Costs and Remaining Budget}
While executing mission $M_i(\tau)$, vehicle $k:=\theta_{M_{i}(\tau)}$ incurs a total offloading cost $C_{i}(\tau)$ for all offloaded tasks $j \in \mathcal{J}_{i}$, given by
\begin{align}
    C_i(\tau) 
    = \sum_{j=1}^{|\mathcal{J}_{i}|} C_{i,j}(\tau)
    = \sum_{j=1}^{|\mathcal{J}_{i}|} c_{o} (d^{\text{comm}}_{i,j} +d^{\text{comp}}_{i,j})
\end{align}
where $C_{i,j}(\tau)$ represents the cost of offloading task $j \in \mathcal{J}_i$
and $c_{o}$ is the per-unit cost charged by serve $o \in \mathcal{S}$.

Each vehicle $\theta_{M_{i}(\tau)}$ is allocated a budget $B_i(\tau)$ to complete mission $M_i(\tau)$, the total offloading cost must not exceed its allocated budget $B_i(\tau)$, such as
\begin{equation} \label{cons:MissionBudgetLimit}
        B_i^{\text{rema}}(\tau) = B_i(\tau) - C_i(\tau) \geq 0, \quad \forall M_{i}(\tau) \in \mathbf{M}(\tau) 
\end{equation}  
where $B_i^{\text{rema}}(\tau)$ is the remaining budget of vehicle $\theta_{M_{i}(\tau)}$ after executing mission $M_i(\tau)$.

\subsection{The Optimization Problem}
Finally, the problem of optimizing mission assignment and task offloading in $\mathtt{Oranits}$, with the objective of maximizing the number of missions completed before their deadlines $T_{i}\left(\tau\right)$, can be formulated as follows:
\begin{subequations} \label{prob:P1-Global}
\begin{align}
\hspace{-0.2cm} 
\mathscr{P}_{1}:\ \underset{ 
    \mathbf{D}\left(\tau\right)
}{\max} ~ 
    & \sum_{M_{i}(\tau) \in \mathbf{M} (\tau)}
    \mathds{1}_{\left\{ \delta_{i} \left(\tau\right)\leq T_{i}\left(\tau\right)\right\} }
     \label{prob:P1-Global_a} \\
    \mathrm{s.t.}~\, 
    & 
    \eqref{cons:MissionVehicleLimit},\,
    \eqref{cons:MissionVehicle},\,
    \eqref{cons:MissionOnlyOneOrder},\,
    \eqref{cons:MissionOrderUnique},\,
    \eqref{cons:MissionOrderBeforeAfter},\,\,
    \eqref{cons:MissionCompletionTimeBound},\,
    \eqref{cons:MissionBudgetLimit}.\label{prob:P1-Global_b}
\end{align}
\end{subequations}

\begin{proposition}
    Problem~$\mathscr{P}_{1}$ in \eqref{prob:P1-Global} is NP-hard.
\end{proposition}
\begin{proof}
   We establish the NP-hardness of problem~$\mathscr{P}_{1}$ via a reduction. Consider a set of $Z$ missions $\mathcal{M}(\tau) = \{M_1, M_2, \cdots, M_Z\}$, where the objective is to determine a scheduling order for their execution on a vehicle $k$, such that the maximum number of missions are completed within the given time frame $T_{i}(\tau) = \tau$. This problem is analogous to the classical scheduling problem with deadlines on a single machine, where tasks must be scheduled to maximize the number of completed tasks within their deadlines. It is well known that this problem is NP-hard in scheduling theory \cite{lawler1994knapsack}. Thus, problem~$\mathscr{P}_{1}$ is also NP-hard.
\end{proof}

\section{Metaheuristic Approach}
\label{sec:Metaheuristic-Approach}
We introduce a metaheuristic approach to solving problem $\mathscr{P}_{1}$ at the cloud, inspired by the ARO algorithm \cite{wang2022artificial}. ARO is modeled on the natural behaviors of wild rabbits, incorporating three main ideas: 
\begin{itemize}
    \item \textbf{Exploration}: Rabbits forage far from their nests to reduce the risk of predator encounters.
    \item \textbf{Exploitation}: They dig multiple burrows near the nest and randomly choose one as shelter, enhancing their chances of evading predators.
    \item \textbf{Transition}: The shift from exploration to exploitation simulates the rabbits' rapid escape response, reflective of their vulnerable position in the food chain.
\end{itemize}
These behaviors are mathematically encoded in ARO, enabling strong performance across a wide range of applications. However, the original ARO suffers from premature convergence, limited global search capability, and sensitivity to the initial population distribution \cite{wang2024improved}, which reduces its effectiveness in complex and multimodal problems. To address these issues, we propose the Chaotic Gaussian-based Global ARO (CGG-ARO) algorithm, a new variant that enhances the initialization, exploration, and exploitation stages through a chaotic mapping process, Gaussian-based population diversification, and a best-solution-guided search direction.

At first, to apply the metaheuristic algorithm to problem  $\mathscr{P}_{1}$, let $\mathbf{x}^{g}_p$ denote a feasible solution at generation $g$ for population member $p$. Each solution is composed of two parts $\mathbf{x}^{g}_p = \{\mathscr{M}_{p}^g \cup \mathscr{V}_{p}^g\}$, where (1) $\mathscr{M}_{p}^g$ is the mission index permutation, and (2) $\mathscr{V}_{p}^g$ is a random vehicle index vector with $|\mathscr{M}_{p}^g| = |\mathscr{V}_{p}^g| = Z$, with each vehicle index being appeared exactly $\left\lceil Z/K^* \right\rceil$ times. Given $\mathbf{x}^{g}_p$, the optimal solution of problem $\mathscr{P}_{1}$ can be found as
$\mathbf{D}_{n,:}(\tau) = \big[\langle\mathscr{V}_{p}^g(m), \sum_{n=1}^m{\mathds{1}_{\{\mathscr{V}_{p}^g(n)=\mathscr{V}_{p}^g(m)\}}}\rangle \mid m \in [1,Z], \text{ sorted by } \mathscr{M}_{p}^g \big]
$.

In the subsequent sections, we present the proposed CGG-ARO algorithm in detail, highlighting the enhancement techniques introduced into each of the three stages of the original ARO algorithm. We first describe the improved initialization stage.

\subsection{Improved initialization stage}
In the ARO algorithm, the initialization phase relies on a randomly generated initial population within the defined search space vector $\{\mathbf{.}\}^{\operatorname{lb}}$ and $\{\mathbf{.}\}^{\operatorname{ub}}$. 
Particularly, the conditions $1\leq\mathscr{M}_p^g(l)\leq Z$ and $1\leq\mathscr{V}_p^g(l)\leq K^*$ should be satisfied for any $l \in [1,|\mathscr{M}^g_p|]$, and thus,
$
\mathscr{V}^{\operatorname{lb}} = \{1\}_{Z},\quad 
\mathscr{V}^{\operatorname{ub}} = \{K^*\}_{Z},\quad 
\mathscr{M}^{\operatorname{lb}} = \{1\}_{Z},\quad 
\mathscr{M}^{\operatorname{ub}} = \{Z\}_{Z}
$, $\mathbf{x}^{\operatorname{lb}} = \{\mathscr{M}^{\operatorname{lb}} \bigcup \mathscr{V}^{\operatorname{lb}}\}$, and $\mathbf{x}^{\operatorname{ub}} = \{\mathscr{M}^{\operatorname{ub}} \bigcup \mathscr{V}^{\operatorname{ub}}\}$. 
However, this conventional approach often lacks sufficient diversity in the initial population, risking premature convergence or getting trapped in local optima, especially in complex landscapes with many local minima.

However, this conventional random initialization often fails to provide sufficient population diversity, which may lead to premature convergence or entrapment in local optima \cite{nguyen2020new}, particularly in complex search landscapes with numerous local minima.

\begin{align}\label{eq:chaotic_aro}
x^{g+1}_{p}(l) =
    \begin{cases} 
    \frac{x^g_p(l)}{\rho}, & 0 \leq x^g_p(l) < \rho; \\
    \frac{x^g_p(l)- \rho}{0.5 - \rho}, & \rho \leq x^g_p(l) < 0.5; \\
    \frac{1 - \rho - x^g_p(l)}{0.5 - \rho}, & 0.5 \leq x^g_p(l) < 1 - \rho; \\
    \frac{1 - x^g_p(l)}{\rho}, & 1 - \rho \leq x^g_p(l) < 1.
    \end{cases}
\end{align}
To address this issue, we incorporate the Piecewise Chaotic Map (PCM) into the initialization stage. PCM \cite{gao2022review} generates chaotic sequences through a deterministic process and offers two main advantages for optimization. First, it improves the uniformity and diversity of the initial population, which is essential for effective global search in evolutionary algorithms. Second, compared with conventional random initialization and many commonly used chaotic maps, PCM distributes candidate solutions more evenly over the search space. This broader coverage helps the algorithm avoid early stagnation and improves its potential to approach the global optimum. Specifically, we first generate all individuals $\mathbf{x}^g_p$ (e.g., $g = 0$) randomly and then refine them using the initialization scheme in \cite{gao2022review}. Each element $x^{g+1}_p(l)$ with $l \in [1,|\mathbf{x}^g_p|]$ is updated as in Eq. \eqref{eq:chaotic_aro}, where $\rho\in(0,\,0.5)$.

\subsection{Improved Detour Foraging Stage}
\label{sec:improved_detour}
In the original ARO algorithm, the \textit{Detour Foraging Strategy} is intended to drive exploration. However, this mechanism mixes exploration and exploitation within a single update process by incorporating a randomly selected individual, thereby biasing the search toward existing solutions \cite{wang2022artificial}. As a result, search diversity is reduced and global exploration is weakened, which increases the risk of premature convergence. To overcome this limitation, CGG-ARO introduces a modified detour foraging strategy that explicitly separates exploration and exploitation into two distinct update modes, thereby improving both search diversity and optimization efficiency.

During the exploration stage, the movement of each individual is driven by a stochastic Gaussian process \cite{isamura2023metaheuristic} with adaptive noise scaling. Unlike the original mechanism, this update does not depend on any specific individual in the population; instead, it adjusts the search step according to the overall population diversity. The update rule is given by:
\begin{equation}
    \mathbf{x}^{g+1}_p = \mathbf{x}^g_p + \mathbf{r}_1\mathcal{N}(0, \boldsymbol{\sigma})
    \label{eq:cgg_aro_01_exploration}
\end{equation}
where $\mathbf{r}_1$ is a binary random vector with elements from $\{0,1\}$ controlling dimensional updates, and $\boldsymbol{\sigma}$ is the per-dimension standard deviation computed across the current population:
\begin{equation}
    \boldsymbol{\sigma} = \operatorname{std}(\mathbf{x}_1^g, \mathbf{x}_2^g, \cdots, \mathbf{x}_{P}^g)
    \label{eq:cgg_aro_std}
\end{equation}
where $P$ denotes the number of individuals. This formulation yields adaptive exploratory steps: when the population is highly diverse, larger perturbations are generated to encourage broader search; as the population gradually converges, the perturbation magnitude decreases naturally, allowing a smooth transition toward exploitation.

\begin{figure}[h]
    \centering
\includegraphics[width=0.7\columnwidth]{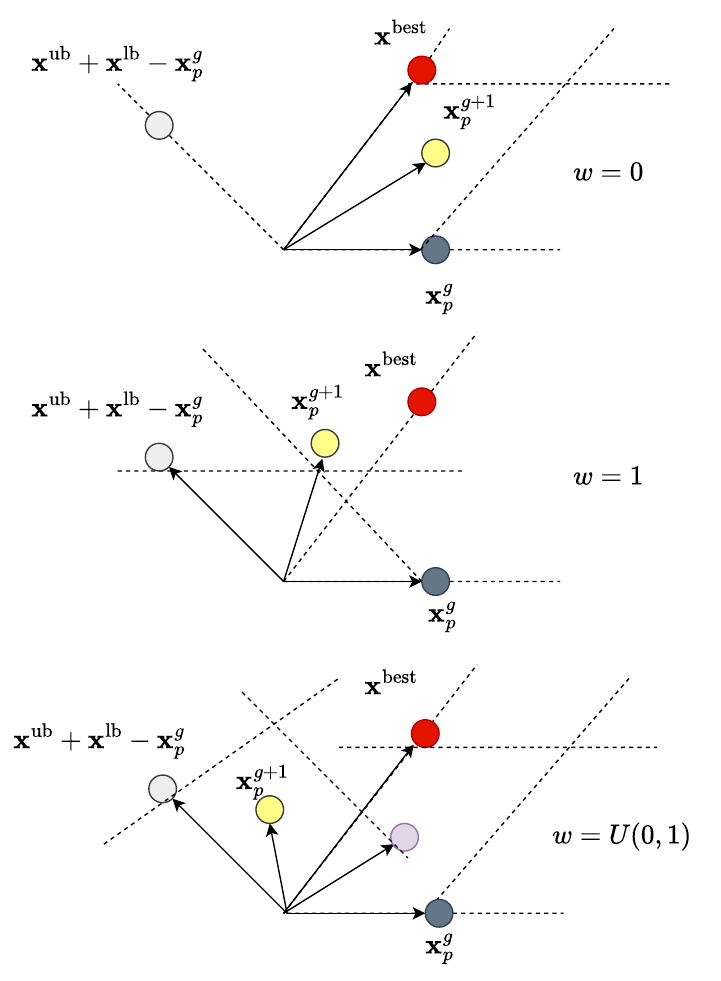}
    \caption{The opposition-based global operator.}
    \label{fig:cgg_aro_phase2}
\end{figure}

In the exploitation stage, the search focuses on refining candidate solutions around promising regions by combining an opposition-based learning term with guidance from the global best solution $\mathbf{x}^{\operatorname{best}}$. The position update is defined as:
\begin{equation}
    \mathbf{x}^{g+1}_p = \mathbf{x}^g_p + \mathbf{r}_2 \big[ w\mathbf{d}_1 + (1 - w)\mathbf{d}_2\big]
    \label{eq:cgg_aro_02_exploitation}
\end{equation}
where $w$ is a probabilistic weighting factor determined by a uniformly distributed random variable $r$, such that:
\begin{equation}
    w = 
    \begin{cases} 
        0, & r < 0.2; \\
        1, & 0.2 \leq r < 0.8; \\
        U(0,1), & r \geq 0.8
    \end{cases}\
    \label{eq:cgg_aro_02_exploitation_w}
\end{equation}
and $\mathbf{r}_2$
is a binary random vector with elements from $\{0,1\},$ and
$\mathbf{d}_1 = \mathbf{x}^{\operatorname{ub}} + \mathbf{x}^{\operatorname{lb}} - \mathbf{x}^g_p
\label{eq:obl}$ represents the opposition-based learning component \cite{nguyen2020new}. This encourages exploration of both the current solution and its opposite. The complementary term
$\mathbf{d}_2 = \mathbf{x}^{\operatorname{best}} - \mathbf{x}^g_p
$ directs the search toward the global best solution, thus balancing exploration and exploitation. This mechanism is illustrated in \autoref{fig:cgg_aro_phase2}.

\subsection{Improved Random Hiding Stage}
In the original ARO algorithm, the \textit{Random Hiding Stage} is designed for exploitation, with the aim of refining candidate solutions in promising regions. However, it relies solely on random vectors and does not exploit historical information such as the global best solution or interactions with peer individuals. Consequently, it often behaves more like an exploratory mechanism than a true exploitation strategy, which may lead to slow convergence and a higher probability of stagnation in local optima.
To address this limitation, we introduce an enhanced \textit{Random Hiding Strategy} that incorporates both exploitation capability and population diversity. The improved mechanism adopts two update scenarios to balance convergence and randomness. The primary update rule combines guidance from the global best solution $\mathbf{x}^{\operatorname{best}}$ with information from randomly selected individuals:
\begin{equation}
    \mathbf{x}^{g+1}_p = \mathbf{x}^{g}_{p'} + (2U(0,1) - 1) \mathbf{r} (\mathbf{x}^{\operatorname{best}} - \mathbf{x}^{g}_{p''})
    \label{eq:cgg_aro_03}
\end{equation}
where $p'$ and $p''$ are randomly selected individual indexes ($p \ne p' \ne p''$). $U(0,1)$ is a uniformly distributed random number, and $\mathbf{r}$ is a vector represents the rabbit's running characteristic \cite{wang2022artificial}. This update rule promotes targeted refinement around promising regions while preserving stochastic diversity in the population.

\begin{algorithm}[t]
\footnotesize
\caption{CGG-ARO Algorithm}
\label{algo:CGG-aro}
\begin{algorithmic}[1]
\State {\bf Input:} Population size $P$, maximum number of iterations $g_{\max}$, lower bound $\mathbf{x}^{\operatorname{lb}}$, upper bound $\mathbf{x}^{\operatorname{ub}}$.
\State {\bf Initialization:} Randomly generate $[\mathbf{x}_1^0, \mathbf{x}_2^0, \cdots, \mathbf{x}_P^0]$ within $[\mathbf{x}^{\operatorname{lb}}, \mathbf{x}^{\operatorname{ub}}]$, then refine them using the PCM-based rule in Eq.~\eqref{eq:chaotic_aro};
\State Set $g = 0$;
\State Evaluate the fitness for the population and determine the best solution $\mathbf{x}^{\operatorname{best}}$;
\While {$g < g_{\operatorname{max}}$}
    \State Compute the population standard deviation using Eq.~\eqref{eq:cgg_aro_std};
    \For {$p \in [1:P]$}
        \State Calculate the energy factor $A$ \cite{wang2022artificial};    
        \If{$A > 1$}
            \If{$U(0,1) > 0.5$}
                \State Calculate the solution $\mathbf{x}^{g+1}_p$ using Eq.~\eqref{eq:cgg_aro_01_exploration};
            \Else
                \State Calculate the solution $\mathbf{x}^{g+1}_p$ using Eq.~\eqref{eq:cgg_aro_02_exploitation};
            \EndIf
        \Else
            \If{$U(0,1) > 0.5$}
                \State Calculate the solution $\mathbf{x}^{g+1}_p$ using Eq.~\eqref{eq:cgg_aro_03};
            \Else
                \State Apply ARO's original updating rule;
            \EndIf
        \EndIf
        \State Retain the better solution based on fitness value;
    \EndFor
    \State Update the best solution $\mathbf{x}^{\operatorname{best}}$;
    \State Set $g = g + 1$;
\EndWhile
\State {\bf Return}: $\mathbf{x}^{\operatorname{best}}$.
\end{algorithmic}
\end{algorithm}

\subsection{Algorithmic Procedure and Complexity Analysis}
Overall, the proposed CGG-ARO improves exploration through PCM-based initialization and Gaussian-based stochastic search, while exploitation is strengthened by the best-solution-guided opposition mechanism and the enhanced random hiding strategy. These components complement each other to provide a better balance between global search diversity and local refinement, thereby reducing the risk of premature convergence and improving optimization robustness. The general procedure of the proposed CGG-ARO algorithm is summarized in \autoref{algo:CGG-aro}, and its computational complexity is analyzed below.

Let $D = |\mathbf{x}_p^g|$ denote the solution dimension and $T_f$ denote the fitness evaluation cost for a single solution. During initialization, generating and refining the population requires $\mathcal{O}(PD)$, while evaluating the initial population requires $\mathcal{O}(PT_f)$. Therefore, the initialization cost is $\mathcal{O}(P(D+T_f))$.
In each iteration, computing the population standard deviation requires $\mathcal{O}(PD)$, updating all individuals requires $\mathcal{O}(PD)$, and evaluating all candidate solutions requires $\mathcal{O}(PT_f)$. Therefore, the per-iteration complexity is $\mathcal{O}(P(D+T_f))$. Over $g_{\max}$ iterations, the total worst-case time complexity is
\begin{align}
    \mathcal{O}\big(g_{\max}P(D+T_f)\big).
\end{align}

In practice, the runtime is typically dominated by the fitness evaluation term $T_f$, especially when the objective function involves simulation, machine learning model training, mission dependencies, or scheduling constraints. In such cases, the overall complexity can be approximated as $\mathcal{O}\big(g_{\max}PT_f\big)$.

\begin{figure}[h]
    \centering
    \begin{subfigure}{0.49\columnwidth}
        \centering
        \includegraphics[width=\textwidth]{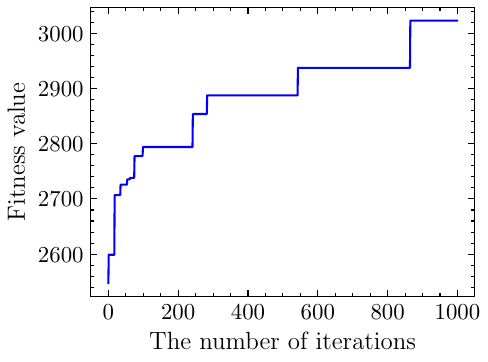}
        \label{fig:system_benefits}
    \end{subfigure}
    \hfill
    \begin{subfigure}{0.49\columnwidth}
        \centering
        \includegraphics[width=\textwidth]{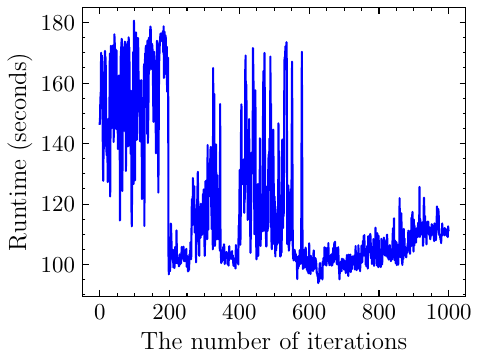}
        \label{fig:running_time}
    \end{subfigure}
    \caption{Objective value (left side) and running time (right side) of the proposed CGG-ARO algorithm.}
    \label{fig:time_perf}
\end{figure}

\section{Deep Reinforcement Learning Approach}
\label{sec:DRL-Approach}
Although the CGG-ARO algorithm is effective for optimization tasks, its performance is often hindered by high computational overhead due to its inherently iterative nature. As illustrated in \autoref{fig:time_perf}, this method typically requires around $1000$ iterations, with each iteration taking between $100$ and $180$ seconds, resulting in a total runtime of nearly $2300$ minutes. Such intensive computational demands render them impractical for real-time deployment in dynamic Open RAN-based ITS environments.
\begin{figure}[h]  
    \centering  
    \includegraphics[width=0.85\columnwidth]{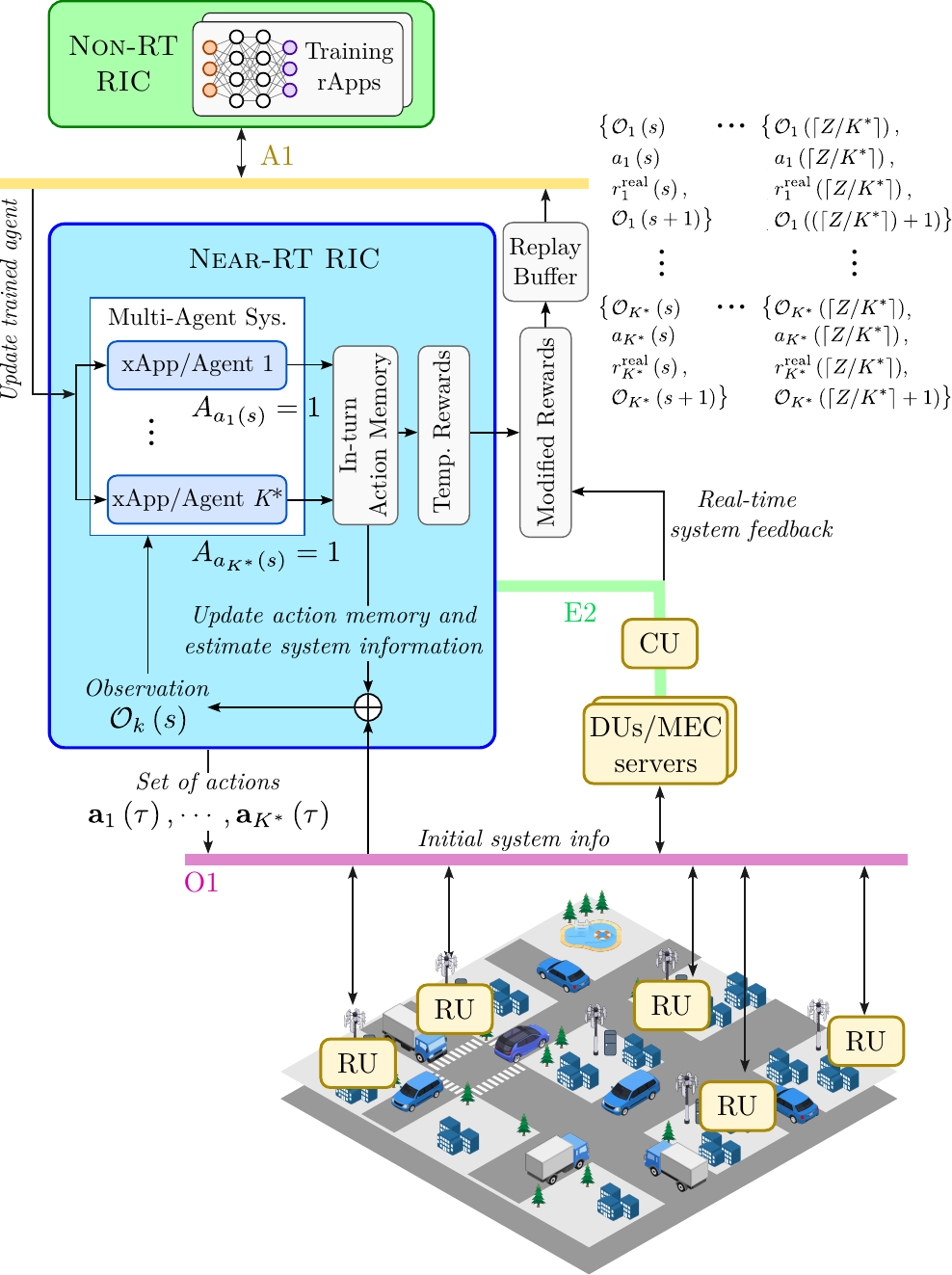} 
    \caption{The proposed multi-agent DRL framework within the Open RAN-based ITS.}  
    \label{fig:deeplearning_mc}  
\end{figure}  

To overcome this limitation, we propose a DRL-based approach leveraging a double deep Q-learning network (DDQN) due to its suitability for discrete action spaces and its ability to mitigate overestimation bias, providing a stable and efficient learning framework for mission assignment and task offloading in dynamic ITS environments. In this framework, $K^*$ agents are associated with $K^*$ vehicles and initially trained within the Non-RT RIC of the Open RAN architecture, as illustrated in \autoref{fig:deeplearning_mc}. These agents are then fed to Near-RT RIC via the A1 interface for real-time decision-making, enabling adaptive and decentralized task allocation. Agents share a common replay buffer to facilitate cooperative learning and utilize real-time system feedback to inform updates. Missions are sequentially assigned and evaluated both in simulation and real-world scenarios. During real-world execution, delayed reward signals, tied to mission outcomes, limit the immediacy of feedback. To handle this, agents record their state-action-reward transitions in the shared replay buffer, supporting off-policy learning via mini-batch training to minimize sample correlation and enhance learning efficiency.

Upon completion of mission assignments, each agent executes its selected actions, and the resulting total system profit is distributed evenly among participating vehicles. To penalize conflicting or infeasible decisions, negative rewards are assigned, reinforcing cooperative behavior. This multi-agent learning strategy, based on MA-DDQN, enables scalable, real-time, and efficient task distribution in complex ITS environments.
In the $\mathtt{Oranits}$ framework, the design, training, and inference of DRL are structured as follows:
\begin{itemize}
    \item \textbf{Environment modeling:} A high-fidelity simulation replicates vehicle mobility, wireless channel dynamics, and network topology. The Open RAN architecture, including open interfaces (\textit{e.g.} E2, A1 \cite{tang2023ai}) and controllers (Non-RT RIC, Near-RT RIC), is integrated to enable realistic policy deployment and state feedback.

    \item \textbf{Offline training at Non-RT RIC:} The DRL model is trained offline at the non-real-time RIC using historical or simulated datasets.
    
    \item \textbf{Model deployment to Near-RT RIC:} Once converged, the trained policy is compressed and deployed as an xApp to the near-RT RIC for online inference. The inference module operates within sub-second constraints, making real-time mission assignment decisions based on current state observations. Post-mission feedback is collected to inform future updates or retraining.
    
    \item \textbf{Periodic policy update:} The near-RT RIC gathers runtime data and system statistics, periodically reporting to the non-RT RIC. This supports policy refinement and continuous learning to adapt to changing network dynamics and traffic patterns.
\end{itemize}

\subsection{Agents}
Each agent observes dynamic information in the environment, including road conditions, vehicle positions, incoming mission requests, and the status of MEC servers. Agents then make real-time decisions regarding task assignment and scheduling, which are then communicated to their corresponding vehicles.
At initialization, agents receive mission data from the nearest RU, along with contextual updates from nearby vehicles and the network. The decision-making process proceeds iteratively until a termination condition is met, typically signaled by reward feedback indicating that all missions have been processed or that no valid selections remain.

After each selection cycle, agents record their experience, including the current state, selected action, received reward, and next state, into a shared replay buffer. This allows for more informed decision-making in subsequent iterations. While each agent follows an independent policy, they share exploration data via this buffer, fostering collaborative learning and improving overall system performance.

\subsection{Observations}
We define the agents' observations at decision step $s$ using several key components that represent the system state.

\textit{Mission assignment status}: Let $\mathbf{A}(s) = \{A_1(s), A_2(s), \cdots, A_Z(s)\}$ denote the mission assignment memory, where $A_k(s) \in \{0, 1\}$ for all $k \in \{1, 2, \cdots, Z\}$. A value of $A_k(s) = 1$ indicates that mission $k$ has already been selected by a vehicle. The decision step $s$ ranges from $0$ to $\lceil Z/K^* \rceil$, where $Z$ is the total number of missions and $K^*$ is the number of agents (or vehicles). Assigning a mission $M_k(\tau)$ to a different vehicle after it has been selected is considered invalid. This vector $\mathbf{A}(s)$ is updated continuously as agents make selections.

\textit{Road conditions}: We denote the road state information at time $s$ as $\mathbf{R}(s) = \{\mathbf{R}_1(s), \mathbf{R}_2(s), \cdots, \mathbf{R}_{|\mathbf{R}(s)|}(s)\}$. Here, each $\mathbf{R}_i(s) = \{R_i(s), {d}_i(s)\}$, with $i\in [1, |\mathbf{R}(s)|]$, represents the state of road segment $i$, where: $R_i(s) \in {0, 1, 2, 3, 4}$ indicates the road status: free flow, stable, slow, congested, or severely congested (as defined in Section~\ref{sec:problem_descriptions}); and $d_i(s) \in \{0, 1\}$ reflects offloading availability: $1$ indicates that the MEC/cloud server is under-loaded and can accept tasks, and $0$ indicates that it is overloaded.

\textit{Vehicle status}: Vehicle information, one of the most critical observation components, provides insight into the current location and status of each vehicle. Let $\mathbf{V}(s) = {\mathbf{V}_1(s), \mathbf{V}_2(s), \cdots, \mathbf{V}_{K^*}(s)}$ denote the set of all vehicle states at time $s$, where each $\mathbf{V}_i(s)$ is defined as:
\begin{equation}
    \mathbf{V}_i(s) = \langle \mathbf{P}_i(s), \mathbf{D}_i(s), L_i(s), v_{i, \operatorname{max}} \rangle.
\end{equation}
Herein, $\mathbf{P}_i(s) = \{x_i(s), y_i(s)\}$ are the vehicle's Cartesian coordinates, $\mathbf{D}_i(s)$ denotes the dependency structure of its currently assigned missions, $L_i(s)$ indicates the number of missions assigned to vehicle $i$, and $v_{i,\text{max}}$ is the maximum allowable speed of vehicle $i$. 

\textit{Mission information}:
Let $\mathbf{J}(s) = \{\mathbf{J}_1(s),\mathbf{J}_2(s),\cdots, $ $\mathbf{J}_{M(\tau)}(s)\}$ represent the set of mission details. Each mission $\mathbf{J}_i(s)$ is represented as:
\begin{equation}
    \mathbf{J}_i(s) =  
    \langle 
        r_{i}^{\text{start}}, 
        r_{i}^{\text{end}}, 
        r_{i}^{\star}, 
        \mathcal{M}^{-}_{i}, 
        \mathcal{M}^{+}_{i}
    \rangle
\end{equation}
where $r_i^{\text{start}}$ and $r_i^{\text{end}}$ are the start and end coordinates of the mission, $r_i^{\star}$ is the best route (in terms of adjusted travel time) between the start and end points, calculated using Dijkstra’s algorithm based on real-time traffic conditions, and $\mathcal{M}_i^-$ and $\mathcal{M}_i^+$ are the sets of predecessor and successor missions, respectively. Note that the best route $r_i^{\star}$ may not always correspond to the shortest geographical distance. Instead, it reflects traffic-adjusted efficiency considering congestion levels on each road segment.

Finally, the complete observation available to agent $k$ at time $s$ is given by: $\mathcal{O}_k(s) = \{\mathbf{R}(s), \mathbf{V}(s), \mathbf{A}(s), \mathbf{J}(s)\}$. Before making any selection, each agent must refresh its local copy of the system state $\mathcal{O}_k(s)$ to ensure decision accuracy. This is especially important as observations evolve rapidly due to ongoing selections and environmental changes.

\subsection{Actions}
During each decision step, the reinforcement learning (RL) agent selects an action $a$ from the action space $\mathcal{A} = \{1, 2, \ldots, Z\}$, where each element corresponds to the index of a mission. The selection process follows an $\varepsilon$-greedy strategy, defined as:
\begin{equation}
     a_k(s) =
    \begin{cases}
        \operatorname{random}(a \in \mathcal{A}), & \text{w/ proba. } \varepsilon \\
        \underset{a \in \mathcal{A}}{\operatorname{argmax}} ~ Q_k(\mathcal{O}_k(s), a), & \text{w/ proba. } 1 - \varepsilon
    \end{cases},
\end{equation}
where $Q_k(\mathcal{O}_k(s), a)$ represents the estimated Q-value for agent $k$ at state $\mathcal{O}_k(s)$ and action $a$.

Let $\boldsymbol{a}_k(\tau) = \{a_k(1), a_k(2), \cdots, a_k(\lceil Z/K^* \rceil)\}$ be the sequence of actions executed by agent $k$ over the assignment period $\tau$. Each element $a_k$ in this sequence identifies the index of a selected vehicle, with its position in the sequence corresponding to the mission index it is assigned to.

The action-value function (Q-function) is updated using the Bellman equation:
\begin{IEEEeqnarray}{rCl}
    Q_k\big(\mathcal{O}_k(s), a_k(s)\big) &=& r_k(s) + \gamma \max_{a' \in \mathcal{A}} Q_k\big(\mathcal{O}_k(s+1), a'\big) \qquad\\
    V\big(\mathcal{O}_k(s)\big) &=& \max_{a \in \mathcal{A}} ~ Q_k\big(\mathcal{O}_k(s), a\big)
\end{IEEEeqnarray}
where $r_k(s)$ is the immediate reward received after taking action $a_k(s)$, $\gamma \in [0,1]$ is the discount factor, and $a'$ represents the optimal action in the subsequent state.

\subsection{Rewards}
The reward function for an agent’s action $a_k(s)$ aims to reflect the real-world impact of mission outcomes and task dependencies. While a simple success-based reward may capture mission completion, it fails to account for queuing order, dependencies, and total vehicle benefit, which are often only available after full mission execution.

To better align with system-wide goals, we define a composite reward that incorporates multiple factors:
\begin{itemize}
    \item \textbf{Actual mission performance:} Agents are rewarded based on the completion status and profitability of their assigned missions, including shared benefits across all executed tasks.

    \item \textbf{Penalty for failed or infeasible missions:} Negative rewards are assigned if an agent selects a mission already taken or one that fails due to constraints.

    \item \textbf{Dependency-aware incentives:} Agents receive bonus rewards for resolving mission dependencies that unblock subsequent tasks. However, delays result in reduced rewards.
\end{itemize}

The adjusted reward function is given by:
\begin{align} \label{eq:reward}
    r^{\text{real}}_k(s) &= \mathds{1}_{\{\delta_{a_k(s)} \leq \tau\}}\mathds{1}_{\{A_{a_k(s)}(s-1) \neq 1\}} \notag \\
    &\times \big(\Gamma_1M^{bc}_{a_k(s)}  + \Gamma_2 B^{\text{rema}}_{a_k(s)} + r_k^{\text{share}}(\tau) + \Gamma_3r^{\text{dep}}_k(s)\big) \notag \\
    & - \mathds{1}_{\{A_{a_k(s)}(s-1) = 1\}} \big(\Gamma_4M^{bc}_{a_k(s)}  + \Gamma_5 B_{a_k(s)}(\tau)\big)
\end{align}
where $M^{bc}_{a_k(s)}$ is the benefit coefficient of mission $M_{a_k(s)}$, and $\Gamma_{\{1, \cdots, 5\}}$ are tunable balancing weights. In \eqref{eq:reward}, the first term represents the actual mission performance, the second term accounts for the effect of dependency removal and queue waiting, and the final term denotes the penalty for failed or infeasible missions. The indicator functions $\mathds{1}_{{\cdot}}$ ensure rewards are only applied when conditions are satisfied (\textit{e.g.} mission is unassigned and completed before the deadline).

Furthermore, the dependency-aware reward term $r^{\text{dep}}k(s)$ is defined as:
\begin{align}
r^{\text{dep}}_k(s) = \big(\lceil Z/K^* \rceil-s\big) 
\big(
    \vert \mathcal{M}_{a_k(s)}^{+} \vert  - \vert \mathcal{M}_{a_k(s)}^{-} \vert + 1
\big)
\end{align}
which captures the reward potential for reducing future constraints through early execution.

Lastly, the shared reward accumulated by agent $k$ from all its completed missions during period $\tau$ is expressed as: 
\begin{align}
    r_k^{\text{share}}(\tau) = \frac{1}{|\sigma_k|}\sum_{m = 1}^{|\sigma_k|} \big(\Gamma_1M^{bc}_{\sigma_k(m)}  + \Gamma_2 B^{\text{rema}}_{\sigma_k(m)}\big)
\end{align}
where $\sigma_k$ denotes the execution order of missions handled by agent $k$.

\section{Numerical Simulation and Discussions}
\label{sec:numerical_simulation}
\subsection{System Parameters}
\begin{figure}[h]
    \centering
    \includegraphics[width=0.7\columnwidth]{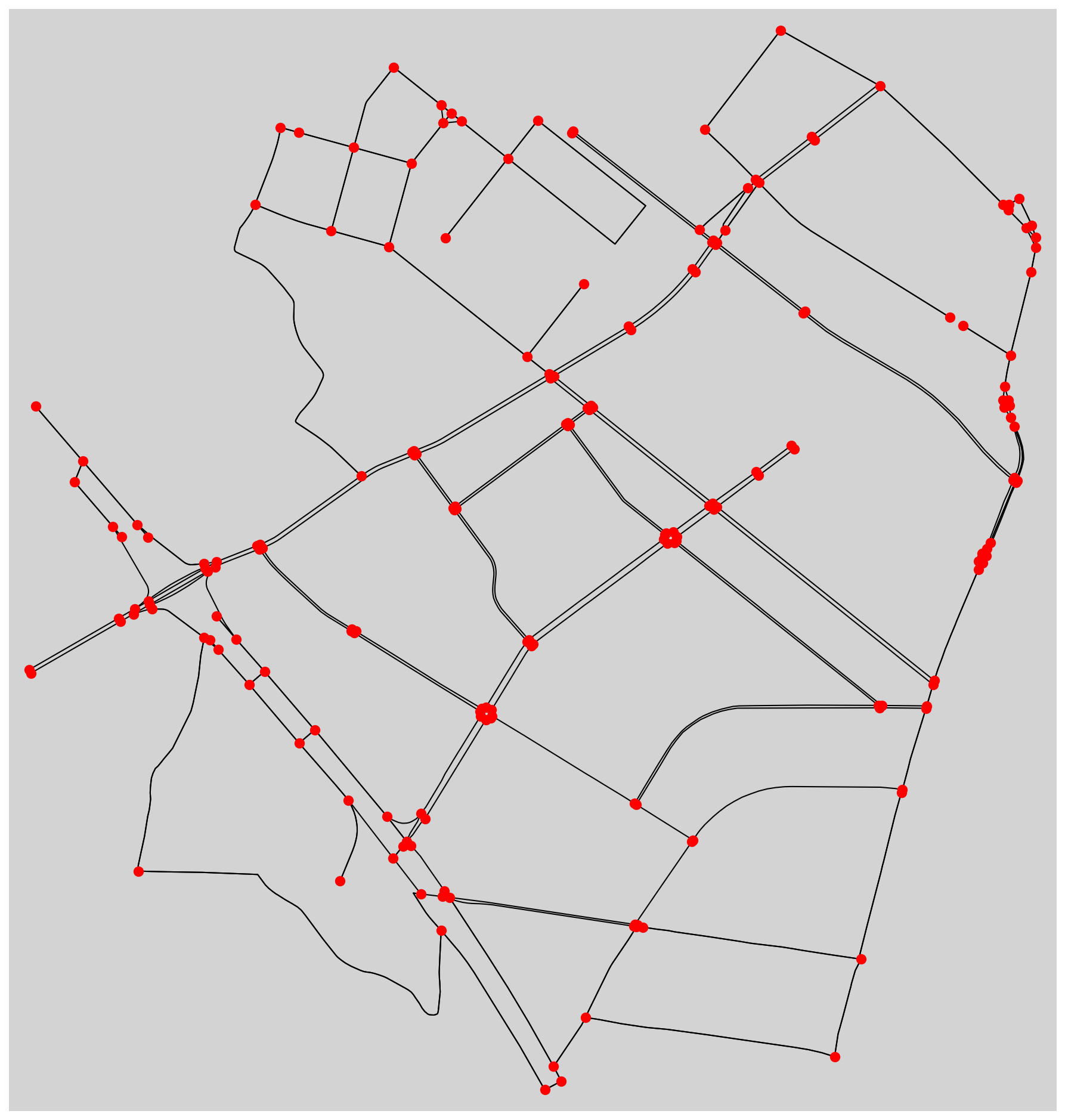}
    \caption{The considered area at the VinUniversity campus, Hanoi, Vietnam, with red intersection points.}
    \label{fig:hnmap}
\end{figure}

As shown in \autoref{fig:hnmap}, we consider a simulation area located at the VinUniversity campus measuring $5000\times 5000$ m\textsuperscript{2}, centered at coordinates $(20.995417, 105.950051)$. This area features a well-developed transportation infrastructure, comprising major roads and intersections, making it a suitable testbed for evaluating the proposed framework.

The simulation involves $K^* = 5$ vehicles tasked with completing up to $Z = 25$ missions within a $\tau = 60$-minute scheduling window. All vehicles are assumed to be within communication range of $|\mathcal{S}^{m}| = 20$ MEC servers and one centralized cloud server ($|\mathcal{S}^{c}| = 1$). Each vehicle achieves the best communication performance within a radius of 100 meters. Additionally, vehicles gain rewards for completed missions proportional to the mission length, specifically $0.025 \times \text{mission length}$. The maximum speed for all vehicles is set to $20$ m/s. The full list of system and model parameters used in our experiments is summarized in \autoref{tab:system-model-parameters}. All simulations are executed on a high-performance computing workstation hosted at the HPC Center of VinUniversity. The system is equipped with an Intel Xeon Gold 6242 CPU (16 cores, 32 threads), $256$~GB RAM, and an NVIDIA RTX A5000 GPU. The simulation environment is implemented in Python~3.10.14, utilizing standard scientific computing libraries such as NumPy, SciPy, and Pytorch.
To ensure statistical significance and account for the inherent stochasticity of meta-heuristic algorithms, each method is executed 15 times on the same mission set, each time with a different random seed. The DRL model is trained on different sets of missions and an evolving transportation system over a period.

\begin{table}[t!]
\caption{System and Model Parameters.}
\label{tab:system-model-parameters}
\centering
\footnotesize
\begin{tabular}{lll}
\toprule
\textbf{Category} & \textbf{Parameter} & \textbf{Value} \\
\midrule
\multirow{9}{*}{System}  
 & Cellular bandwidth & $10$ MHz \\
 & Transmission power & $199.526$ mW \\
 & RU's antennas & $16$ \\
 & Path loss exponent & $3$ \\
 & Number of channels (RUs) & $10$ \\
 & Noise power spectrum density $\sigma^2$ & $-174$ (dBm/Hz) \\
 & Fiber optic transmission rate & $150$ (Gbps) \\
 & Number of vehicles & 5 \\
 & Number of missions & 25 \\
\midrule
\multirow{6}{*}{DRL}  
 & Activation functions & SELU, ELU \\
 & Discount factor $\gamma$ & $0.95$ \\
 & Learning rate & $1 \times 10^{-5}$ \\
 & $\varepsilon$ (initial, decay, min) & $1.0$, $0.99$, $0.05$ \\
 & Batch size & $512$ \\
 & Replay memory size & $10^7$ \\
\midrule
\multirow{3}{*}{Metaheuristic}  
 & Number of iterations & 1000 \\
 & Number of populations & 30 \\
 & Number of different seeds & 15\\
\bottomrule
\end{tabular}
\end{table}
\textbf{Baselines}: To demonstrate the effectiveness of our proposed algorithms, we compare their performance against several state-of-the-art metaheuristic baselines. In addition to well-known techniques such as ARO, SHADE, and L-SHADE, we also include two recently proposed optimization algorithms in our evaluation: 1) \textbf{Equilibrium Optimizer (EO)} \cite{faramarzi2020equilibrium}:  Based on the dynamic equilibrium behavior of mass balance systems, this algorithm guides candidate solutions toward an optimal equilibrium point, efficiently exploring the search space while maintaining rapid convergence. 2) \textbf{Artificial Protozoa Optimizer (APO)} \cite{wang2024artificial}: Inspired by the movement and adaptive behavior of protozoa in nutrient search, this algorithm employs simple yet effective biological strategies to navigate complex optimization landscapes.

To ensure a fair comparison, all algorithms are executed under consistent settings. The parameter configurations for all metaheuristic algorithms are summarized in \autoref{tab:system-model-parameters}. Additionally:
\begin{itemize}
    \item APO uses a neighbor pair value of $np = 2$ and a maximum proportion fraction of $pf_{\text{max}} = 0.1$.
    \item SHADE and L-SHADE are initialized with a weighting factor $\mu_f = 0.5$ and a crossover probability $\mu_{cr} = 0.5$.
    \item EO, ARO, and our proposed CGG-ARO do not require additional parameter tuning.
\end{itemize}
Each scenario is evaluated across multiple trials using different random seeds, and the performance metrics are reported as averages over these runs. This evaluation methodology ensures a robust and fair comparison of optimization performance across diverse simulation scenarios.

\subsection{Performance Evaluation for CGG-ARO}
\begin{table}[H]
    \centering
    \caption{Mean and Standard Deviation Values of Fitness, Completed Missions, and Total Benefits over $15$ Runs.}
    \begin{tabular}{lccc}
        \toprule
        \textbf{Model} & \textbf{Fitness} & \textbf{Completed missions} & \textbf{Total benefits} \\
        \cmidrule[0.4pt](lr{0.12em}){1-1}%
        \cmidrule[0.4pt](lr{0.12em}){2-2}%
        \cmidrule[0.4pt](lr{0.12em}){3-3}%
        \cmidrule[0.4pt](lr{0.12em}){4-4}%
        APO     & $2834.6 {\scriptstyle\pm 51.5}$ 
                & $22.8 {\scriptstyle\pm 0.6}$ 
                & $1138.2 {\scriptstyle\pm 30.9}$ \\
        SHADE   & $2876.8 {\scriptstyle\pm 68.4}$ 
                & $23.2 {\scriptstyle\pm 0.8}$ 
                & $1158.2 {\scriptstyle\pm 39.4}$ \\
        L-SHADE & $2885.6 {\scriptstyle\pm 68.8}$ 
                & $23.4 {\scriptstyle\pm 0.9}$ 
                & $1168.2 {\scriptstyle\pm 43.4}$ \\
        EO      & $2868.3 {\scriptstyle\pm 48.2}$ 
                & $23.3 {\scriptstyle\pm 0.7}$ 
                & $1164.8 {\scriptstyle\pm 35.6}$ \\
        ARO     & $2909.0 {\scriptstyle\pm 76.1}$ 
                & $23.7 {\scriptstyle\pm 0.9}$
                & $1181.5 {\scriptstyle\pm 45.4}$ \\
        \textbf{CGG-ARO} & $\mathbf{2941.2} {\scriptstyle\pm \mathbf{84.6}}$ 
                & $\mathbf{24.0} {\scriptstyle\pm \mathbf{1.0}}$
                & $\mathbf{1198.2} {\scriptstyle\pm \mathbf{49.4}}$ \\
        \bottomrule
    \end{tabular}
    \label{tab:mean_std_cgg_aro_01}
\end{table}
\autoref{tab:mean_std_cgg_aro_01} reports the mean (average) performance and standard deviations of all compared algorithms in terms of fitness, completed missions, and total benefits. 
The proposed CGG-ARO algorithm achieves the best average fitness value ($2941.24$), significantly outperforming ARO ($2909.03$) and L-SHADE ($2885.64$). This improvement is a direct consequence of the modifications introduced in the algorithm, particularly the enhanced initialization using the Piecewise Chaotic Map in \eqref{eq:chaotic_aro}, which provides a more uniformly distributed starting population and facilitates better early-stage exploration.
In terms of mission execution efficiency, CGG-ARO achieves the highest number of completed missions ($24.00$) and total benefits ($1198.16$). This aligns with the structured exploration–exploitation mechanism embedded in CGG-ARO. Specifically, the Gaussian-based exploration phase \eqref{eq:cgg_aro_01_exploration} introduces stochastic diversity proportional to the current population variance, allowing the algorithm to explore broader regions of the search space. Meanwhile, the exploitation phase (Section \ref{sec:improved_detour}), driven by a hybrid of opposition-based learning and global best guidance in \eqref{eq:cgg_aro_02_exploitation} supports rapid refinement in promising areas.
Although CGG-ARO exhibits slightly higher variability in fitness scores (standard deviation = $84.6$) compared to ARO ($76.1$) and L-SHADE ($68.8$), this is expected due to its stronger exploration component. The probabilistic switching weight $w$ in \eqref{eq:cgg_aro_02_exploitation} allows dynamic adjustment between aggressive exploration and focused exploitation, which contributes to both solution diversity and adaptability across different mission scenarios.

\begin{figure}[t]
    \centering
    \begin{subfigure}[b]{0.49\columnwidth}
        \centering
        \includegraphics[width=\linewidth]{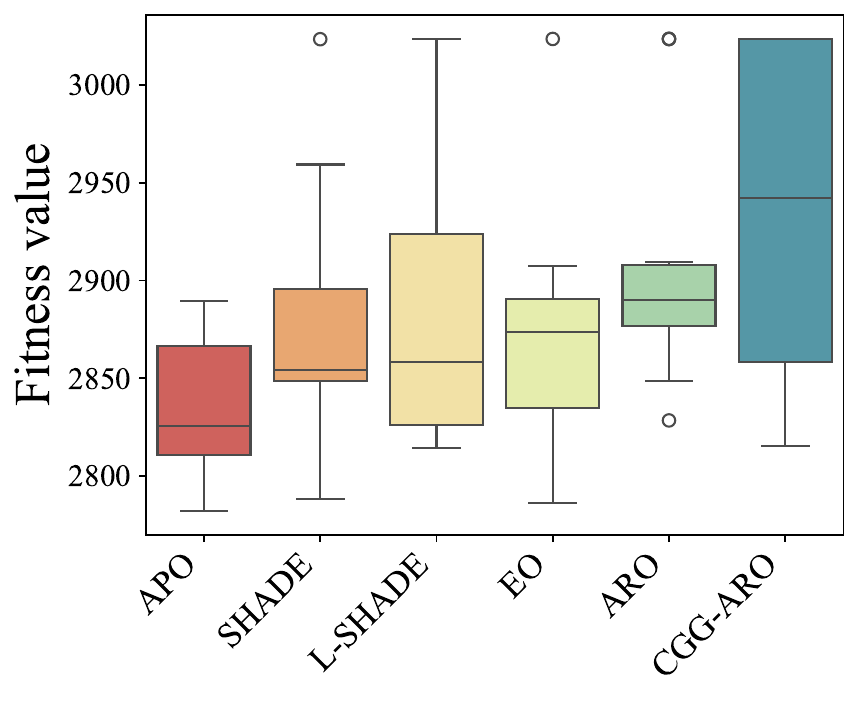}
        \caption{Fitness value.}
        \label{fig:cgg_aro_meta_01}
    \end{subfigure}
    \hfill
    \begin{subfigure}[b]{0.49\columnwidth}
        \centering
        \includegraphics[width=\linewidth]{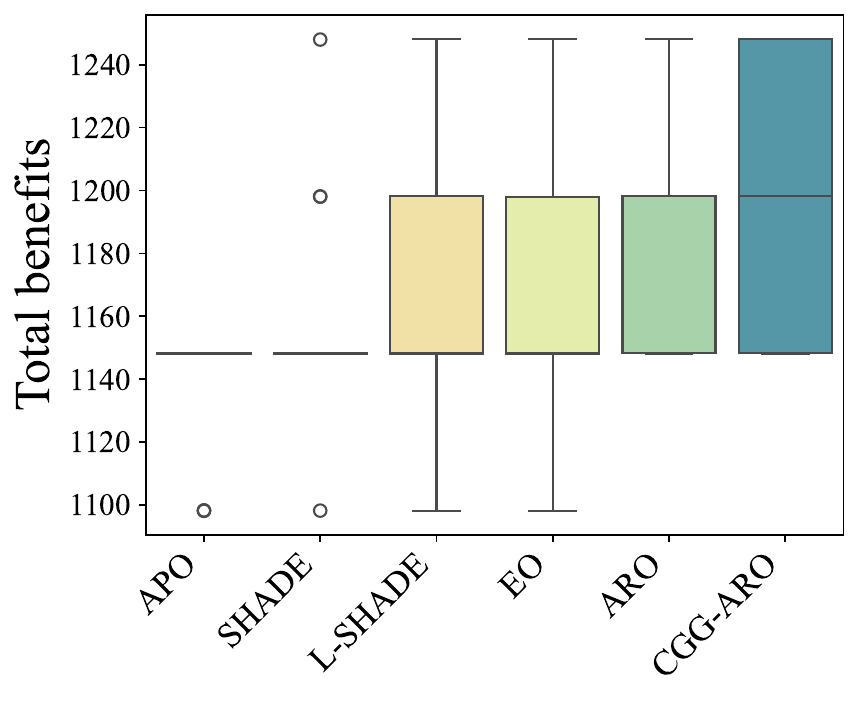}
        \caption{Total benefit.}
        \label{fig:cgg_aro_meta_02}
    \end{subfigure}

    \caption{Boxplot comparison of (a) the fitness value and (b) the total benefits of the compared algorithms.}
    \label{fig:cgg_aro_meta_boxplot}
\end{figure}

The boxplots in \autoref{fig:cgg_aro_meta_boxplot} provide further insights into distributional behavior. CGG-ARO not only achieves higher median fitness and total benefits but also maintains a relatively compact interquartile range, suggesting stable performance across trials. The reduced number of outliers compared to standard ARO implies that CGG-ARO is less prone to erratic behavior, a result attributed to the improved exploitation strategy in \eqref{eq:cgg_aro_03} that incorporates both global best and randomly selected agents to refine the search locally.

\begin{figure*}[t]
    \centering
    \begin{subfigure}[b]{0.32\textwidth}
        \centering
        \includegraphics[width=\linewidth]{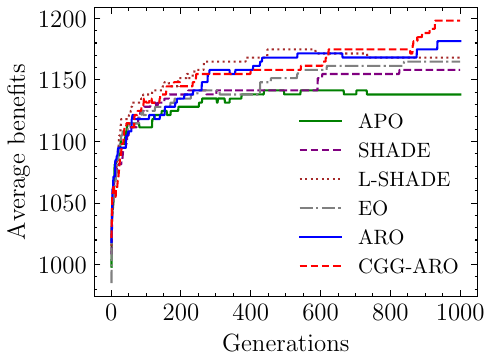}
        \caption{Average benefits.}
        \label{fig:cgg_aro_meta_conv_benefits}
    \end{subfigure}
    \hfill
    \begin{subfigure}[b]{0.32\textwidth}
        \centering
        \includegraphics[width=\linewidth]{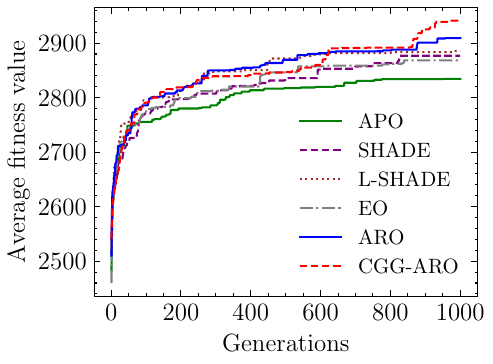}
        \caption{Average fitness.}
        \label{fig:cgg_aro_meta_conv_fitness}
    \end{subfigure}
    \hfill
    \begin{subfigure}[b]{0.305\textwidth}
        \centering
        \includegraphics[width=\linewidth]{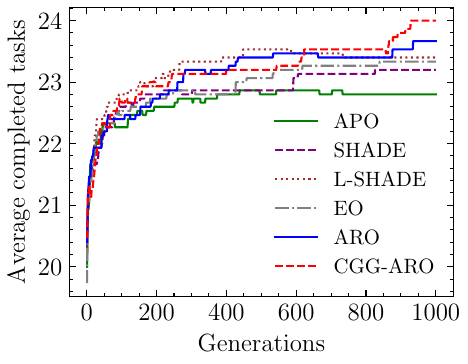}
        \caption{Completed missions.}
        \label{fig:cgg_aro_meta_conv_tasks}
    \end{subfigure}

    \caption{Convergence behavior of the compared algorithms across different performance metrics.}
    \label{fig:Fig.8}
\end{figure*}

Furthermore, the convergence curves in \autoref{fig:Fig.8} confirm that CGG-ARO consistently converges faster than other algorithms. Its early-stage improvement is largely due to the diversity introduced by chaotic initialization and Gaussian perturbation, while the later-stage stability is ensured by the exploitation mechanisms that rely on historical knowledge.

In summary, the improvements in CGG-ARO are not merely empirical but are analytically supported by its underlying mathematical design. The combination of chaotic sequence generation, Gaussian-guided exploration, and adaptive exploitation contributes to its strong ability to balance global and local search. This balance is particularly crucial in complex environments such as \texttt{Oranits}, where task dependencies, resource constraints, and dynamic mission timing must be optimized simultaneously.

\subsection{DDQN with and without Modified Network}
\begin{figure}[htb]
    \centering
    \includegraphics[width=0.8\columnwidth]{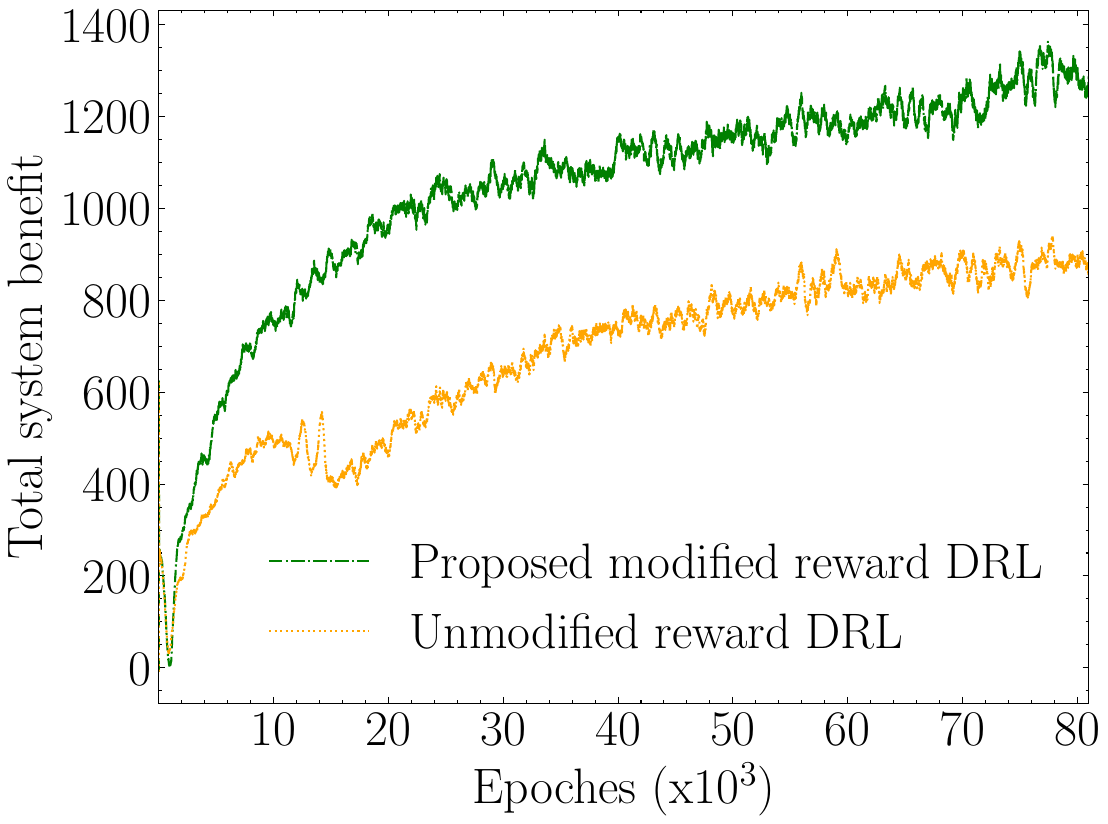}
    \caption{The convergence of MA-DDQN.}
    \label{fig:reward}
\end{figure}

\noindent{}\textbf{Convergence of MA-DDQN}:
Fig.~\ref{fig:reward} illustrates the performance of the MA-DDQN system under two reward schemes: the proposed scheme, which incorporates a modified reward based on system feedback after the completion of all missions, and the unmodified baseline scheme. The proposed reward scheme achieves a significantly higher total system benefit, stabilizing at approximately $1200$ after $50,000$ training epochs. In contrast, the unmodified scheme converges to a much lower value of around $600$, highlighting its suboptimal performance.
The primary limitation of the unmodified scheme lies in its myopic reward structure, where agents prioritize immediate gains over long-term system outcomes. This shortsightedness hinders the system’s ability to optimize task dependencies and vehicle allocation, ultimately leading to reduced overall performance. In comparison, the reward curve of the proposed scheme is smoother and exhibits better convergence, indicating a more effective learning process for optimizing and stabilizing system behavior.
These results suggest that the proposed reward structure enables a more balanced trade-off between exploration and exploitation, fostering improved coordination among agents. Conversely, the unmodified scheme's higher volatility and slower convergence reflect difficulties in aligning individual agent actions with the broader system objectives.

We now present a comparative performance analysis between the proposed MA-DDQN method and several state-of-the-art metaheuristic algorithms, including APO, L-SHADE, EO, and the proposed CGG-ARO. The evaluation is conducted using $15$ distinct Sets of Missions (SoM), with each algorithm executed over $10$ independent trials per set. Performance is assessed based on the average number of completed missions and the corresponding fitness values.

\begin{table}[htb]
    \centering
    \caption{Average Completed Missions Across Different Mission Sets.}
    \begin{tabular}{cccccc}
        \toprule
        \textbf{SoM} & \textbf{APO} & \textbf{L-SHADE} & \textbf{EO} & \textbf{CGG-ARO} & \textbf{MA-DDQN} \\
        \cmidrule[0.4pt](lr{0.12em}){1-1}%
        \cmidrule[0.4pt](lr{0.12em}){2-2}%
        \cmidrule[0.4pt](lr{0.12em}){3-3}%
        \cmidrule[0.4pt](lr{0.12em}){4-4}%
        \cmidrule[0.4pt](lr{0.12em}){5-5}%
        \cmidrule[0.4pt](lr{0.12em}){6-6}%
        1  & 14.0  & 14.1  & 14.0  & 14.6  & \textbf{15.0} \\
        2  & 21.2  & 21.3  & 21.4  & \textbf{22.3}  & 21.0  \\
        3  & 15.2  & 15.3  & 15.6  & 16.6  & \textbf{19.0} \\
        4  & 19.3  & 18.9  & 18.9  & \textbf{20.4}  & 8.0   \\
        5  & 18.3  & 19.2  & 18.9  & \textbf{19.7}  & 17.0  \\
        6  & 14.8  & 15.3  & 15.4  & 16.4  & \textbf{18.0} \\
        7  & 14.5  & 14.9  & 15.1  & 16.7  & \textbf{20.0} \\
        8  & 18.0  & 18.2  & 18.5  & \textbf{18.7}  & 17.0  \\
        9  & 11.6  & 11.4  & 11.5  & 12.0  & \textbf{22.0} \\
        10 & 10.7  & 10.8  & 10.9  & 11.0  & \textbf{19.0} \\
        11 & 19.4  & 19.3  & 20.1  & \textbf{21.1}  & 18.0  \\
        12 & 16.0  & 17.0  & 17.2  & 18.0  & \textbf{19.0} \\
        13 & 15.7  & 16.6  & 16.0  & 17.5  & \textbf{18.0} \\
        14 & 12.9  & 12.7  & 13.0  & 13.6  & \textbf{19.0} \\
        15 & 18.3  & 18.2  & 17.9  & 18.3  & \textbf{21.0} \\
        \bottomrule
    \end{tabular}
    \label{tab:drl_completed_tasks}
\end{table}

\textbf{Mission completion}: 
Table \ref{tab:drl_completed_tasks} presents the average number of completed missions for each algorithm across different Sets of Missions. The results demonstrate that the proposed MA-DDQN scheme delivers strong performance, achieving the highest number of completed missions in $8$ out of $15$ sets, specifically Sets $1\ (15.0)$; $6\ (18.0)$; $3,\ 10,\ 12,\ 14\ (19.0)$; $7\ (20.0)$; $9\ (22.0)$ outperforming CGG-ARO and other metaheuristic algorithms in these instances. This highlights MA-DDQN's effectiveness in dynamic mission allocation scenarios, likely due to its reinforcement learning approach enabling adaptive decision-making. However, CGG-ARO achieves the highest completion count in Sets $2\ (22.3),\ 4\ (20.4),\ 11\ (21.1),$ and 
$15\ (21.0)$, underscoring the strengths of evolutionary-based methods in structured optimization scenarios where iterative refinement excels. Notably, in Set $4$, CGG-ARO completes $20.4$ missions compared to MA-DDQN’s $8.0$, illustrating a significant performance gap in certain contexts. 

\begin{table}[htb]
    \centering
    \caption{Average Fitness Values Across Different Mission Sets.}
    \begin{tabular}{cccccc}
        \toprule
        \textbf{SoM} & \textbf{APO} & \textbf{L-SHADE} & \textbf{EO} & \textbf{CGG-ARO} & \textbf{MA-DDQN} \\
        \cmidrule[0.4pt](lr{0.12em}){1-1}%
        \cmidrule[0.4pt](lr{0.12em}){2-2}%
        \cmidrule[0.4pt](lr{0.12em}){3-3}%
        \cmidrule[0.4pt](lr{0.12em}){4-4}%
        \cmidrule[0.4pt](lr{0.12em}){5-5}%
        \cmidrule[0.4pt](lr{0.12em}){6-6}%
        1  & 1997.3  & 2052.8  & 2043.0  & 2158.5  & \textbf{2294.3} \\
        2  & 3036.9  & 3066.5  & 3080.1  & 3216.6  & \textbf{3402.6} \\
        3  & 2408.2  & 2445.4  & 2479.0  & 2600.1  & \textbf{3062.6} \\
        4  & 2738.6  & 2725.5  & 2743.3  & \textbf{2931.5}  & 1318.7 \\
        5  & 2575.9  & 2679.0  & 2636.5  & \textbf{2729.4}  & 2348.9 \\
        6  & 2262.4  & 2266.0  & 2305.3  & 2468.8  & \textbf{3056.7} \\
        7  & 2261.4  & 2323.1  & 2334.2  & 2558.0  & \textbf{3103.4} \\
        8  & 2450.7  & 2476.2  & 2477.2  & \textbf{2526.4}  & 2348.7 \\
        9  & 1805.4  & 1787.0  & 1792.4  & 1843.5  & \textbf{3098.0} \\
        10 & 1784.5  & 1796.8  & 1809.2  & 1821.5  & \textbf{2937.7} \\
        11 & 2943.4  & 2935.0  & 3036.4  & \textbf{3168.2}  & 2544.2 \\
        12 & 2390.9  & 2459.6  & 2518.0  & \textbf{2628.7}  & 2572.4 \\
        13 & 2431.9  & 2494.8  & 2449.3  & 2674.5  & \textbf{2716.4} \\
        14 & 1909.1  & 1899.5  & 1912.0  & 2011.1  & \textbf{2892.6} \\
        15 & 2280.1  & 2301.8  & 2234.6  & 2310.7  & \textbf{2783.3} \\
        \bottomrule
    \end{tabular}
    \label{tab:drl_fitness_values}
\end{table}
\textbf{Fitness value}:
Table~\ref{tab:drl_fitness_values} presents the average fitness values across different mission sets, reflecting the optimization quality of each algorithm. MA-DDQN achieves the highest fitness values in 9 out of 15 mission sets (sets $1, 2, 3, 6, 7, 9, 11, 12, 14$), with notable peaks such as 3402.6 (set 2) and 3098.0 (set 9), highlighting its superior performance in dynamic mission allocation tasks. This success can be attributed to MA-DDQN's self-learning approach and modified reward structure, which enable adaptive decision-making in complex, dynamic environments. However, CGG-ARO outperforms MA-DDQN in sets 4 ($2931.5 \text{ vs. } 1318.7$), 5 ($2729.4$ vs. $2348.9$), and 8 ($2526.4$ vs. $2348.7$), indicating its competitive edge in specific scenarios. CGG-ARO’s evolutionary-based approach excels in structured optimization problems, where iterative refinement over generations yields high-quality solutions. In conclusion, while MA-DDQN demonstrates overall dominance in dynamic settings, CGG-ARO's performance in select mission sets underscores the value of evolutionary algorithms in certain structured contexts, suggesting potential for hybrid approaches to leverage the strengths of both methods. By computing the mean values from \autoref{tab:drl_completed_tasks} and \autoref{tab:drl_fitness_values}, we observe that the proposed method achieves notable gains, improving overall system benefits by $12.5\%$ and mission assignment efficiency by $7.7\%$, which demonstrates the effectiveness of the approach.

\section{Conclusion and Future Work}
\label{sec:cons}
This work addressed the problem of mission assignment and task offloading in Open RAN-based ITS, where mission interdependencies and offloading costs are often overlooked. We introduced $\mathtt{Oranits}$, a system model that explicitly incorporates these factors to improve overall system performance through vehicle cooperation. To solve this problem, we proposed two algorithms: a metaheuristic-based evolutionary method for one-slot optimization and a modified reward-based multi-agent DRL framework for dynamic mission allocation.  Extensive simulations showed that $\mathtt{Oranits}$ outperforms existing approaches, achieving improvements of approximately $12.5\%$ in overall system benefit and $7.7\%$ in mission assignment performance. These results confirm the effectiveness of the proposed method in enhancing both system efficiency and task allocation quality.

Future work will focus on integrating federated learning for privacy-preserving adaptation across distributed nodes and combining DRL with graph neural networks to enhance decision-making in complex, interdependent mission scenarios. We will also explore advanced actor–critic methods (e.g., PPO and A3C) to further improve learning stability and scalability in dynamic environments. These directions aim to strengthen the intelligence, adaptability, and scalability of $\mathtt{Oranits}$ for large-scale ITS deployments.

\bibliographystyle{./IEEEtran}
\bibliography{ ./IEEEabrv, refs}

\end{document}